\documentclass{aastex631}
\usepackage{microtype}
\usepackage{multirow}
\usepackage{gensymb}
\usepackage{xcolor}
\usepackage{amsmath}
\usepackage[bottom]{footmisc}
\usepackage{comment}
\usepackage{float}
\usepackage{placeins}
\graphicspath{{./}{lense_thirring/}}
\usepackage{xcolor}
\newcommand{\mbh}{M_{\rm BH}}
\newcommand{\mstar}{M_{\star}}
\newcommand{\msol}{M_{\odot}}

\patchcmd{\endthebibliography}{\newpage}{}{}{}
\begin{document}

\title{Strong Field Tidal Disruptions of Rotating Stars by Spinning Black Holes}

\author{Niles Mondal}
\affiliation{Indian Institute of Technology, Kanpur 208016, India} 

\author{Kowsona Chakraborty}
\affiliation{Indian Institute of Technology, Kanpur 208016, India}

\author{Pratap Alte}
\affiliation{Indian Institute of Technology, Kanpur 208016, India}

\author{Pritam Banerjee}
\affiliation{Indian Institute of Astrophysics, Koramangala, Bangalore 560034, India}

\author{Tapobrata Sarkar}
\affiliation{Indian Institute of Technology, Kanpur 208016, India}

\begin{abstract}
We study tidal disruptions of rotating solar-type stars by a rapidly spinning supermassive
black hole with general-relativistic smoothed-particle hydrodynamics. 
To capture strong-field effects, we place the pericentre at a few gravitational radii of a black hole, and 
compare with weak-field encounters at a larger pericentre around a lighter black hole.
We vary the black hole spin, the orbital inclination and the stellar
rotation, using rigidly and differentially rotating stars with spin aligned, anti-aligned or
perpendicular to the orbital angular momentum.
In the strong field, the black hole spin decides whether a core
survives on equatorial and inclined orbits, and stellar rotation reduces the surviving core in
every partial disruption. At the larger pericentre, as in Newtonian studies, anti-aligned rotation
instead protects the star from disruption. In both cases the tide reverses an anti-aligned
spin just before pericentre. The outcome differs because in the strong field the tide keeps
acting on the reversed core long after pericentre. Deep encounters leave cores spinning above break-up,
which shed mass and angular momentum as they settle. Full disruptions approach the canonical 
fallback law, whereas the late-time slope of partial disruptions is not universal and is set by the debris energy 
distribution near the surviving core.
\end{abstract}


\section{Introduction} \label{sec:intro}

Tidal disruption events (TDEs) offer a unique theoretical and observational laboratory for probing the gravitational environments of supermassive 
black holes (SMBHs). A TDE occurs when the self-gravity of a star is overwhelmed by the tidal forces of a nearby black hole \citep{Hills1975, FrankRees, Rees}. 
Depending on the orbital impact parameter, these encounters result in either full TDEs, where the star is completely shredded into a stream of debris, or partial 
TDEs, where a self-bound remnant core survives the encounter. Assuming that the returning debris efficiently circularizes and accretes on a viscous 
timescale shorter than the fallback timescale, the resulting accretion luminosity will directly track the mass supply, so that the peak of the observed light 
curve closely follows the mass fallback rate. While the late-time light curve of a full disruption typically declines at a rate proportional to 
$t^{-5/3}$ \citep{Rees, Phinney1989, EvansKochanek1989}, partial encounters introduce complex stream dynamics that are known to 
produce marked deviations from this canonical law \citep{GuillochonRamirezRuiz2013, CoughlinNixon2019}.  

In TDEs, the characteristic scale of the encounter is commonly taken as the tidal radius, estimated
via Newtonian methods as $r_{\rm t}\simeq R_\star\left(\mbh/\mstar\right)^{1/3}$, at which the tidal field of the BH
overcomes the self-gravity of the star, while the strength of the encounter is measured by the
impact parameter $\beta=r_{\rm t}/r_{\rm p}$, with $r_{\rm p}$ the pericentre distance
\citep{Hills1975, CarterLuminet1983}. Because $r_{\rm t} \sim \mbh^{1/3}$ whereas the
gravitational radius $r_{\rm g}\sim \mbh$, a star is disrupted outside the BH event horizon only for BH
masses $\mbh\lesssim10^{8}\,\msol$ for a solar-type star and a
non-rotating BH. Heavier black holes will instead swallow the star instead of a disruption \citep{Hills1975, Rees}. Introduction of black hole spin alters this bound, since the horizon of a rapidly
rotating Kerr black hole is smaller and prograde orbits can have smaller pericentres compared to retrograde ones
\citep{Kesden2012, Leloudas2016}. 
Typically the orbital plane of an incoming star bears no relation to the spin axis of
the hole, so inclined, off-equatorial encounters should be common. 

Analytic treatments cannot accurately follow the nonlinear hydrodynamics of tidal disruption, 
hence one typically resorts to numerical simulation. Early work used Newtonian smoothed particle
hydrodynamics (SPH) and simple polytropic stars \citep{Nolthenius1982, EvansKochanek1989,
Laguna1993, Ayal2000}. SPH has been refined for over three decades now, and present day simulations 
result in high-resolution surveys of the parameter space \citep{Lodato2009,
GuillochonRamirezRuiz2013, Mainetti2017, LawSmith2020}. Important insights have also resulted from
semi-analytic models of the deformed star \citep{CarterLuminet1983, Kochanek1994} and from
grid-based calculations of the debris flow \citep{ShiokawaEtAl2015, Sadowski2016}. 
In our previous works \citep{GarainSarkar2024, GarainSarkar2025, Mahapatra2026a, Mahapatra2026b, Mahapatra2026c},
we explored TDEs involving both solitary and binary white dwarfs in intermediate mass Kerr BH backgrounds. 
In these works, the pericenter distance of the centre of mass of the star or the binary was taken to be 
$\gtrsim 20\,r_{\rm g}$. A hybrid approach was followed there, with the stellar particles following exact geodesics
and self gravity being handled in a Newtonian fashion. As argued in these works, this was sufficiently robust 
in the regimes of interest considered there. 

For the deep, relativistic encounters which will be our interest here, however, Newtonian gravity is inadequate: at
$r_{\rm p}\sim{\rm few}\times r_{\rm g}$ the orbital dynamics, the tidal tensor and the geometry of the debris
stream are all controlled by strong-field effects. This has motivated a hierarchy of relativistic
treatments, from generalised Newtonian potentials that reproduce Kerr geodesics
\citep{Tejeda2017}, through relativistic hydrodynamics on a fixed Kerr background
\citep{ChengEvans2013, Ryu2020I, Ryu2020II, Ryu2020III, Ryu2020IV}, to fully general-relativistic
SPH \citep{LiptaiPrice2019, Liptai2019} and general-relativistic magnetohydrodynamic simulations
of the resulting accretion flow \citep{Andalman2022}. 

Two strong-field effects dominate such encounters. The first is strong apsidal precession
\citep{Hayasaki2013, Dai2015, Bonnerot2016, ShiokawaEtAl2015}, and the second is frame dragging. 
In particular, a spinning BH drags inertial frames around it, so an orbit inclined to the equatorial plane undergoes
Lense-Thirring nodal precession. For TDEs, this results in the debris stream precessing out of its original
plane, so that self-intersection is delayed or suppressed altogether, and the disc that eventually forms
may precess as a rigid body or tear into misaligned rings \citep{StoneLoeb2012, Franchini2016,
Hayasaki2016, Liptai2019}. Spin and inclination also modify the disruption itself, that is, 
the tidal tensor evaluated along an inclined Kerr geodesic differs from
its Newtonian counterpart in both magnitude and orientation. This results in prograde and retrograde orbits (at
fixed $\beta$) encountering different tidal fields, and the outcome of a marginal encounter can
switch between full disruption, partial disruption and no disruption at all as the inclination is
varied \citep{Kesden2012, Gafton2015, GaftonRosswog2019, Mahapatra2026a}.
Despite substantial progress in relativistic TDE modelling, the combined effects of black hole spin, orbital inclination and stellar rotation on strongly relativistic partial disruptions remain poorly understood. Recent first-principles population modelling shows that partial disruptions contribute approximately 30\% of detected events in flux-limited samples and are essential for reproducing the observed TDE population \citep{Piran2026}. Understanding their dynamics is therefore important for connecting the physics of individual disruptions to observable TDE demographics.
Existing studies have investigated BH spin and stellar rotation predominantly in weak-field encounters, while strong-field simulations have focused mainly on non-spinning BHs and full disruptions. Consequently, the interplay between relativistic tidal forces, frame dragging and stellar rotation near the threshold of full disruption remains largely unexplored. Previous studies of partial disruptions around $\sim10^6\,\msol$ BHs found that BH spin has a relatively modest influence on fallback rates and surviving core masses, although it can determine whether a deeply plunging star survives \citep{Tejeda2017, GaftonRosswog2019}. Newtonian simulations of rotating stars have demonstrated that prograde stellar spin enhances tidal stripping, whereas retrograde spin suppresses it \citep{Golightly2019, Kagaya2019, SacchiLodato2019} at the same $r_{\rm p}$. Stellar spin can also influence the evolution of repeating partial TDEs \citep{Bandopadhyay2026}. 

Strong-field simulations have predominantly considered non-spinning BHs. Relativistic hydrodynamic calculations have explored partial disruptions of realistic main-sequence stars in Schwarzschild spacetime \citep{Ryu2020I, Ryu2020II, Ryu2020III, Ryu2020IV}, while studies of full disruptions have examined relativistic debris dynamics and accretion flows \citep{Andalman2022, Chan2026}. However, extreme encounters near the direct-capture limit and rapidly spinning BHs with oblique spin axes remain insufficiently explored. Existing investigations of tilted encounters around spinning BHs have likewise focused on full disruptions \citep{Price2024}. \citet{Calderon2026} investigated the effects of relativistic stream precession on TDE light curves using radiation-hydrodynamics simulations, finding that precession can delay the luminosity peak for highly inclined encounters around massive BHs. 

A further unresolved question concerns the role of the stellar rotation profile. Existing rotating-star TDE simulations have assumed rigid rotation, leaving it unknown whether differential rotation modifies the response to relativistic tidal fields independently of the total stellar angular momentum. Recent studies  have further highlighted the importance of differential rotation in TDEs, which can produce distinct observational signatures, such as late-time rebrightening flares and secondary peaks in the fallback rate \citep[e.g., GSN;][]{Wang2024GSN}. Although the latter study considers the disruption of a common-envelope progenitor rather than an isolated stellar progenitor, it further motivates investigating the role of the rotation profile in isolated stellar progenitors. This work addresses these gaps by systematically investigating strongly relativistic partial TDEs around spinning BHs, examining how BH spin, orbital inclination and stellar rotation jointly determine disruption outcomes and remnant properties. By considering differentially rotating stellar configurations alongside rigidly rotating models, we further test whether the internal distribution of angular momentum introduces effects that cannot be captured by the total stellar spin alone. Our study thus probes the previously unexplored intersection of strong-field gravity, partial disruption and stellar rotation, focusing on the transition between survival, partial disruption and full disruption.
We consider both prograde and retrograde spins, $a^*=\pm0.98$, and three orbital inclinations (near-polar, $45^\circ$ and equatorial), and with the pericentre deep in the relativistic regime, we follow the disrupted star through pericentre and beyond in order to extract the bound core mass, the debris energy distribution, the fallback rate and its slope, the rotation of the surviving core. We performed the simulations using our in-house GRSPH code \citep{Banerjee2026a}, which adopts the general relativistic hydrodynamic formulation of \citet{LiptaiPrice2019}, accurately captures the gravity of the Kerr BH and incorporates our recently developed improved stellar self-gravity treatment \citep{Banerjee2026b} necessary for simulating near-horizon TDEs.

The paper is organised as follows. In Section~\ref{sec:methods} we describe the simulation setup: the black hole, stellar and orbital parameters, the construction of inclined parabolic Kerr geodesics from the turning-point structure of the radial potential, the relaxation of the polytropic stellar model, and the rotation laws used for the rotating stellar models. In Section~\ref{sec:pp} we analyse the centre-of-mass geodesics as test particle orbits. The strong-field results are presented in Section~\ref{sec:results} and compared with weak-field runs at $r_{\rm p}\simeq25\,r_{\rm g}$ in Section~\ref{sec:25rg}, and we conclude in Section~\ref{sec:conclusion}.

\section{Methodology and simulation setup}
\label{sec:methods}

We simulate the tidal disruption of a solar-type star by a spinning supermassive black hole (SMBH) using general-relativistic smoothed-particle hydrodynamics (GRSPH), in which the star's internal hydrodynamics is evolved on a fixed Kerr background while its centre of mass follows an exact geodesic. In geometrized units, $G=c=1$, so that the gravitational and Schwarzschild radii of the hole are $r_{\rm g}=M_{\rm BH}$ and $r_{\rm s}=2M_{\rm BH}$, respectively.

\subsection{Black hole, star, and orbital parameters}
\label{sec:params}

The black hole has mass $M_{\rm BH}=5\times10^{7}\,M_\odot$ and dimensionless Kerr spin $a^*=\pm0.98$, with the sign convention that $+$ ($-$) denotes prograde (retrograde) rotation relative to the stellar orbital angular momentum. We choose $M_{\rm BH}=5\times10^{7}\,M_\odot$ because it lies close to the Hills mass of a rapidly spinning retrograde black hole for our encounter, a slightly heavier black hole would capture the star whole on that orbit \citep{Hills1975, Kesden2012}. This places every encounter in our set as deep in the strong field as a surviving star can reach for this black hole spin and penetration factor.
The star is a solar-type $\gamma=5/3$ ($n=3/2$) polytrope on a parabolic ($e=1$) orbit with penetration factor, $\beta \equiv \tfrac{r_{\rm t}}{r_{\rm p}}=0.555$, fixing the pericentre at $r_{\rm p}\simeq6.3\,r_{\rm g}$  deep enough into the relativistic regime to probe strong-field effects while remaining outside the marginally bound radius for both spin orientations. We define the orbital inclination $i$ as the angle between the BH spin axis ($z$-axis) and the initial orbital plane, so that $i=90^\circ$ is equatorial and $i\to0^\circ$ (near-polar) is the numerically singular limit of our formalism; together, $(a^*,i)$ span the parameter space explored below. The star is initialized in a parabolic orbit  from $r_0=15\,r_{\rm t}$ with $\phi_0=0$ and $\theta_0=i$, and its initial centre-of-mass velocity is fixed by the Kerr geodesic integrals of motion together with the parabolic turning-point condition, as described next.

\subsection{Geodesic initial conditions}
\label{sec:geodesics}

The trajectory of the stellar centre of mass is initialised as an exact Kerr geodesic in Boyer-Lindquist coordinates $(t,r,\theta,\phi)$, with metric
\begin{equation}
ds^2 = -\left(1-\frac{2Mr}{\Sigma}\right)dt^2
-\frac{4Mar\sin^2\theta}{\Sigma}\,dt\,d\phi
+\frac{\Sigma}{\Delta}\,dr^2
+\Sigma\,d\theta^2
+\left(r^2+a^2+\frac{2Ma^2r\sin^2\theta}{\Sigma}\right)\sin^2\theta\,d\phi^2,
\label{eq:kerr_metric}
\end{equation}
where $M\equiv M_{\rm BH}$, $\Sigma=r^2+a^2\cos^2\theta$, $\Delta=r^2-2Mr+a^2$, and $a=a^*M$. Following \citet{Mahapatra2026a} and \citet{Tejeda2017}, an inclined geodesic is fully specified by the conserved energy $E$, axial angular momentum $L_z$, and Carter constant $Q$. Rather than invoking the special-case (equatorial or spherical) formulas, we determine these directly from the turning-point structure of the radial potential,
\begin{equation}
R(r)=\varepsilon\,(r-r_{\rm p})(r-r_a)(r-r_c)(r-r_d), \qquad \varepsilon\equiv E^2-1,
\end{equation}
which for a parabolic orbit ($\varepsilon=0$, $r_a\to\infty$) reduces to a two-parameter root-finding problem for the remaining turning points $r_c,r_d$:
\begin{subequations}
\begin{align}
f_1(r_c,r_d) &\equiv r_{\rm p} r_c r_d - a^2\,(r_{\rm p}+r_c+r_d)\cos^2 i = 0, \\
f_2(r_c,r_d) &\equiv r_{\rm p} r_c + r_c r_d + r_{\rm p} r_d
- \Big[ r_{\rm s}(r_{\rm p}+r_c+r_d) - 2a\sin i\sqrt{r_{\rm s}(r_{\rm p}+r_c+r_d)} + a^2 \Big] = 0 .
\end{align}
\label{eq:rc_rd_parabolic}
\end{subequations}
Solving Equations~\eqref{eq:rc_rd_parabolic} numerically for $(r_c,r_d)$ at fixed $(r_{\rm p},a^*,i)$ then yields the conserved quantities in closed form,
\begin{equation}
Q = \frac{r_{\rm s}\,r_{\rm p} r_c r_d}{a^2}, \qquad
L^2 = r_{\rm p}(r_c+r_d)+r_c r_d, \qquad
L_z = \sqrt{r_{\rm s}\!\left[(r_{\rm p}+r_c+r_d)-\frac{r_{\rm p} r_c r_d}{a^2}\right]},
\label{eq:conserved_quantities}
\end{equation}
which are substituted into the Kerr first integrals to obtain the stellar centre-of-mass position and velocity at $r_0$, subsequently transformed into the Cartesian-like frame used for the SPH evolution.

\subsection{Stellar model and hydrodynamic evolution}
\label{sec:hydro}
We consider three stellar rotation configurations: non-rotating (NR), rigidly rotating (RR), and differentially rotating (DR). 
The stellar model is relaxed independently of the orbit calculation. For a polytropic equation of state with polytropic index $n=3/2$, we obtain the equilibrium density profile by solving the hydrostatic equilibrium equations. The stellar model is relaxed using established SPH stretch-mapping techniques until the virial ratio, defined as the ratio of kinetic to gravitational potential energy, $(T/|W|)$, satisfies $(T/|W|\lesssim10^{-6})$. The relaxation procedure ensures that residual kinetic motions are negligible compared with the star's gravitational binding energy, indicating that the stellar configuration has reached a sufficiently stable equilibrium.

The RR and DR models are constructed by spinning up an initially relaxed non-rotating SPH star through gradual azimuthal velocity updates. For rigid rotation, the angular velocity is uniform throughout the star, with the target value ($\Omega_{\mathrm{rot}}=\lambda_{\mathrm{rot}}\Omega_{\mathrm{br}}$), where $\Omega_{\mathrm{br}}$ is the stellar break-up angular velocity defined by $\Omega_{\mathrm{br}}=\sqrt{\tfrac{GM_{\star}}{R^{3}_\star}}$ . In contrast, for the differentially rotating stellar model, we adopt the relativistic Komatsu–Eriguchi–Hachisu (KEH) $j$-constant rotation law \citep{Komatsu1989}, whose Newtonian limit corresponds to the Hachisu profile given by :
\begin{equation}
\Omega(s)
=
\frac{\Omega_{\rm rot}}
{1+\left(s/A\right)^2}.
\end{equation}
where  $A$ is the differential rotation parameter that controls the degree of differential rotation and $s$ is the cylindrical distance of stellar element. According to this law the angular velocity of fluid elements inside the star decreases as $s$ increases from $A$ to $R_\star$. The target rotation profile is introduced gradually, while residual motions are damped using artificial viscosity. The star is then evolved for an additional $25$ equilibrium-test interval in units of dynamical time of the RR star, $t_\star=(R_\star^3/GM_\star)^{1/2}$ which is about $1.6\times10^{3}$\,s ($27$\,min) for a solar-type star. This step is performed to verify that the structure remains stable after the relaxation procedure. For DR star this test is performed for additional $40\,t_\star$ to confirm stability.  Rotational equilibrium is assessed by monitoring the mean angular velocity $(\Omega_{\mathrm{avg}}=|\mathbf L|/I)$ and checking its convergence toward the target value. 

To isolate the effect of differential rotation on the energy distribution and subsequent fallback rate from variations in the total rotational energy budget, we carefully matched the global macroscopic properties of RR and DR. For the RR, we adopt a moderate $\lambda_{\mathrm{rot}}=0.3$ (9.3-hour rotation period), pushing beyond the $\lambda_{\mathrm{rot}}=0.2$ regime explored in earlier studies \citep{Golightly2019}. This allows us to construct a hydrodynamically stable configuration without introducing the extreme deformation associated with very rapid rotation. For the DR, we select $\lambda_{\mathrm{rot}}=1.8$ and $A=0.2$ (corresponding to rotation periods of 39.6 hours at the stellar surface and 3.06 hours at $0.2R_\star$ from the centre) by performing a series of equilibrium calculations with different combinations of $\lambda_{\mathrm{rot}}$ and $A$, as the resulting model attains a rotational energy comparable to that of the RR, with $E_{\mathrm{spin}}/E_{\mathrm{grav}}=0.012$. The angular velocity distributions and virial equilibrium tests for the RR and DR stellar models are presented in Appendix~\ref{appendixA}. Stellar spin parameters for both spinning stellar models were chosen such that the relaxed spinning stars retain nearly spherical surfaces. Each simulation was run with $N=5\times10^{5}$ SPH particles and the Wendland $C^4$ kernel \citep{Wendland1995, DehnenAly2012}. The equations of motion and their time integration follow
the GRSPH formulation of \citet{LiptaiPrice2019}. We evolve the stellar fluid using an ideal gas equation of state, with the entropy variable as the evolved thermodynamic quantity in our GRSPH simulations. We adopt the artificial viscosity (AV) shock-capturing criterion of \citet{LiptaiPrice2019}, with AV strength limited by the Balsara switch \citep{Balsara1995}, and a constant viscosity parameter $\alpha_{\rm AV}=1$ ($\alpha_{\rm min}=\alpha_{\rm max}=1$).  We also include artificial conductivity with  $\alpha_u=0.1$ to reduce spurious thermal discontinuities. To establish numerical convergence, we perform additional simulations using $8\times10^{5}$ SPH particles. The corresponding convergence tests are presented in Appendix~\ref{app:convergence}.

\begin{figure}[!tbp]
    \includegraphics[width=\textwidth]{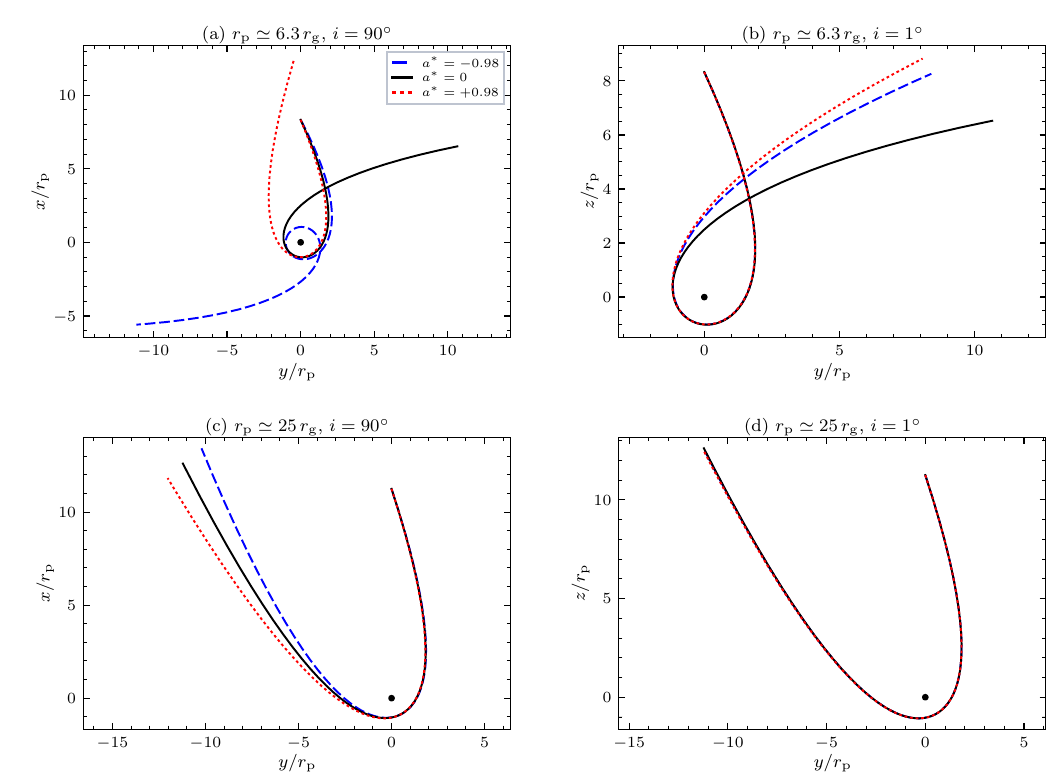}
    \caption{Centre-of-mass geodesics for $M_{\rm BH}=5\times10^7\,M_\odot$, $\beta=0.555$
    (top) and $M_{\rm BH}=4\times10^6\,M_\odot$, $\beta=0.75$ (bottom), from
    $r=15\,r_{\rm t}$ through pericentre and out to $22.5\,r_{\rm t}$, in units of
    $r_{\rm p}$ with the black hole at the origin. Equatorial orbits are shown as $x$
    against $y$, near-polar orbits as $z$ against $y$. Blue dashed, black solid and red
    dotted lines show $a^*=-0.98$, $0$ and $+0.98$. Only the retrograde equatorial orbit in
    panel (a) whirls around the black hole; in panel (d) the three orbits coincide. The launch radius changes neither the turn nor the time inside $2r_{\rm p}$.}
    \label{fig:orbits}
\end{figure}
\section{Relativistic trajectory and duration of the strong tide }
\label{sec:pp}
Before we describe the hydrodynamic results, we study the orbit of the stellar centre of mass as a test particle that follows the parabolic Kerr geodesic of Section~\ref{sec:geodesics}. All stellar models on a given orbit follow the same geodesic, so the orbit sets the duration and strength of the tide but cannot by itself distinguish the spin orientations.

We consider two sets of encounters. The strong-field set has $M_{\rm BH}=5\times 10^7\,M_\odot$ and $\beta=0.555$, which gives the pericentre $r_{\rm p}\simeq6.3\, r_{\rm g}$ (Section~\ref{sec:params}). The weak-field set has $M_{\rm BH}=4\times 10^6\,M_\odot$ and $\beta=0.75$, which gives $r_{\rm p}\simeq 25\, r_{\rm g}$. We use it in Section~\ref{sec:25rg} to separate strong-field effects from the ordinary response of a rotating star. For both sets we integrate the geodesics for $a^*=0,\pm0.98$ on the equatorial ($i=90^\circ$) and near-polar ($i=1^\circ$) orbits. The Schwarzschild orbit does not depend on $i$.

The global coordinate time $t$ is measured in units of the dynamical time of the star. A star responds to changes in the forces acting upon it on its dynamical timescale, which characterizes its ability to restore hydrostatic equilibrium; thus, the number of dynamical times it spends near pericentre measures how much it can respond while the tidal field is strongest.
We call $t_{\rm p}$ the time from the start of a run to
pericentre. The strong-field runs start
at $r_0=15\,r_{\rm t}$, which gives $t_{\rm p}=36.5$--$44.0\,t_\star$ on the geodesics. The
weak-field runs start at $r_0=10\,r_{\rm t}$ ($t_{\rm p}\simeq18.3\,t_\star$).

We call the part of the orbit inside $r=2r_{\rm p}$ the strong-tide phase, and measure from the geodesic how far the orbit turns there, and how long the star stays there, $\Delta t_{2{\rm p}}$. The tidal acceleration across the star roughly falls as $r^{-3}$, so at $2r_{\rm p}$ it is eight times weaker than at the pericentre.  On a Newtonian parabolic orbit
the test body turns through $180^\circ$ inside $2r_{\rm p}$ and
$\Delta t_{2{\rm p}}=(8\sqrt{2}/3)\,\beta^{-3/2}\,t_\star$. This gives $\Delta t_{2{\rm p}}=9.1\,t_\star$ for $\beta=0.555$ and
$\Delta t_{2{\rm p}}=5.8\,t_\star$ for $\beta=0.75$.

Figure~\ref{fig:orbits} shows the geodesics. At $r_{\rm p}\simeq25\,r_{\rm g}$ the orbit (panels c and d)  turns through $192^\circ$--$198^\circ$  inside $2r_{\rm p}$ and the star stays there for $6.3$--$6.5\,t_\star$, about
10 per cent longer than the Newtonian value. For the equatorial case, after the pericentre passage the retrograde orbit stays closer to the black hole than the prograde orbit. For the near-polar case, black hole spin hardly affects the orbit of the star.

At $6.3\,r_{\rm g}$ (panels a and b) the outgoing branches depend strongly on the black hole
spin. In panel (a) the three equatorial orbits leave in very different directions, and in
panel (b) the two Kerr orbits leave close together, apart from the Schwarzschild orbit. On the prograde equatorial orbit, the
Schwarzschild orbit and the near-polar orbits, the orbit turns through
$234^\circ$--$280^\circ$ inside $2r_{\rm p}$, and the star stays there for
$13.0$--$15.1\,t_\star$, $1.4$--$1.7$ times the value from the Newtonian expression of $\Delta t_{2{\rm p}}$. Lense-Thirring precession also tilts the near-polar orbital plane by
about $25^\circ$. This tilt moves the outgoing branch out of the $y$--$z$ plane, in
opposite directions for $a^*=\pm0.98$, so it does not appear in panel (b). The retrograde
equatorial orbit stays longest near the black hole. Its pericentre lies only 8 per cent
outside the marginally bound radius, $r_{\rm mb}\simeq5.8\,r_{\rm g}$, and the star whirls
around the black hole (panel a): the orbit turns through $439^\circ$ inside $2r_{\rm p}$,
and the star stays there for $22.6\,t_\star$, $2.5$ times the value from the Newtonian expression of $\Delta t_{2{\rm p}}$. The geodesic is symmetric about pericentre, so half of $\Delta t_{2{\rm p}}$ comes after it: $3.2\,t_\star$ at $25\,r_{\rm g}$, against $6.5$--$7.6\,t_\star$ on the prograde equatorial, Schwarzschild and near-polar orbits and $11.3\,t_\star$ on the retrograde equatorial orbit at $6.3\,r_{\rm g}$. In the bound-mass and core-spin figures in the following sections, the strong tide therefore ends near $t/t_{\rm p}\simeq1.2$ in both sets ($1.26$ on the retrograde equatorial orbit, $1.10$ for the near-polar NR runs at $25\,r_{\rm g}$).

The two sets differ in $\beta$ as well as in $r_{\rm p}/r_{\rm g}$, so the longer strong
tide at $6.3\,r_{\rm g}$ could come partly from the lower $\beta$ rather than from
relativity. We separate the two effects in Appendix~\ref{app:fixed_beta}.

The point-particle orbits therefore predict a longer strong tide at $6.3\,r_{\rm g}$, with
larger turns inside $2r_{\rm p}$, a tilted orbital plane on the near-polar orbit, and the
longest stay on the retrograde equatorial orbit. How the star responds depends on its
hydrodynamics, which we describe next.

\section{Hydrodynamical evolution and outcomes}
\label{sec:results}

This section presents the strong-field runs ($r_{\rm p}\simeq6.3\,r_{\rm g}$); they are
compared with the runs at $r_{\rm p}\simeq25\,r_{\rm g}$ in Section~\ref{sec:25rg}. Each
encounter is set by the black hole spin $a^*$, the orbital inclination $i$ and the stellar
model. We use $a^*=\pm0.98$ on the equatorial ($i=90^\circ$), inclined ($i=45^\circ$) and
near-polar ($i=1^\circ$) orbits. The Schwarzschild orbit ($a^*=0$) does not depend on
inclination and serves as the reference for all three. Here $i$ is the angle between the
black hole spin axis and the orbital plane, so $i=90^\circ$ is the equatorial orbit. The
star is non-rotating (NR), rigidly rotating (RR) or differentially rotating (DR). For the
rotating stars $\theta_{\rm s}$ is the angle between the stellar spin and the orbital
angular momentum: the star rotates with the orbit for $\theta_{\rm s}=0^\circ$ (aligned),
against it for $\theta_{\rm s}=180^\circ$ (anti-aligned), and its spin lies in the orbital
plane for $\theta_{\rm s}=90^\circ$. All comparisons are made at fixed $a^*$ and $i$. Time
is given as $t/t_{\rm p}$ (Section~\ref{sec:pp}). We
describe the bound core and its spin (Section~\ref{sec:core}), the energy distribution of
the stripped debris (Section~\ref{sec:dmde}) and the fallback rate with its late-time slope
(Section~\ref{sec:fallback}).

\subsection{Bound core mass and core spin}
\label{sec:core}

\begin{figure}[tbp]
    \includegraphics[width=\textwidth]{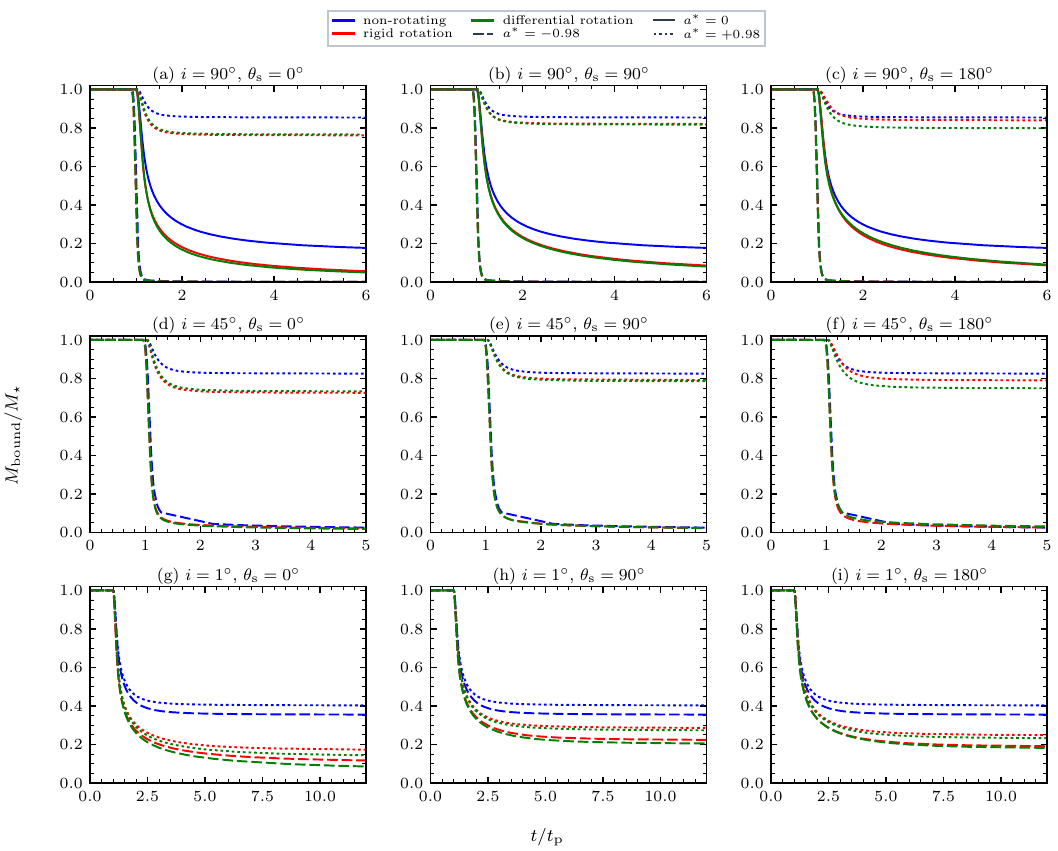}
    \caption{Bound mass fraction $M_{\rm bound}/M_\star$ at $r_{\rm p}\simeq6.3\,r_{\rm g}$.
Rows show the equatorial ($i=90^\circ$), inclined ($i=45^\circ$) and near-polar
($i=1^\circ$) orbits, where $i$ is the angle between the black hole spin axis and the
orbital plane. Columns show the angle $\theta_{\rm s}$ between the stellar spin and the
orbital angular momentum. Blue, red and green lines show the NR, RR and DR models. Solid,
dashed and dotted lines show $a^*=0$, $-0.98$ and $+0.98$.}
    \label{fig:mbound}
\end{figure}

\begin{figure}[tbp]
    \includegraphics[width=1.0\textwidth]{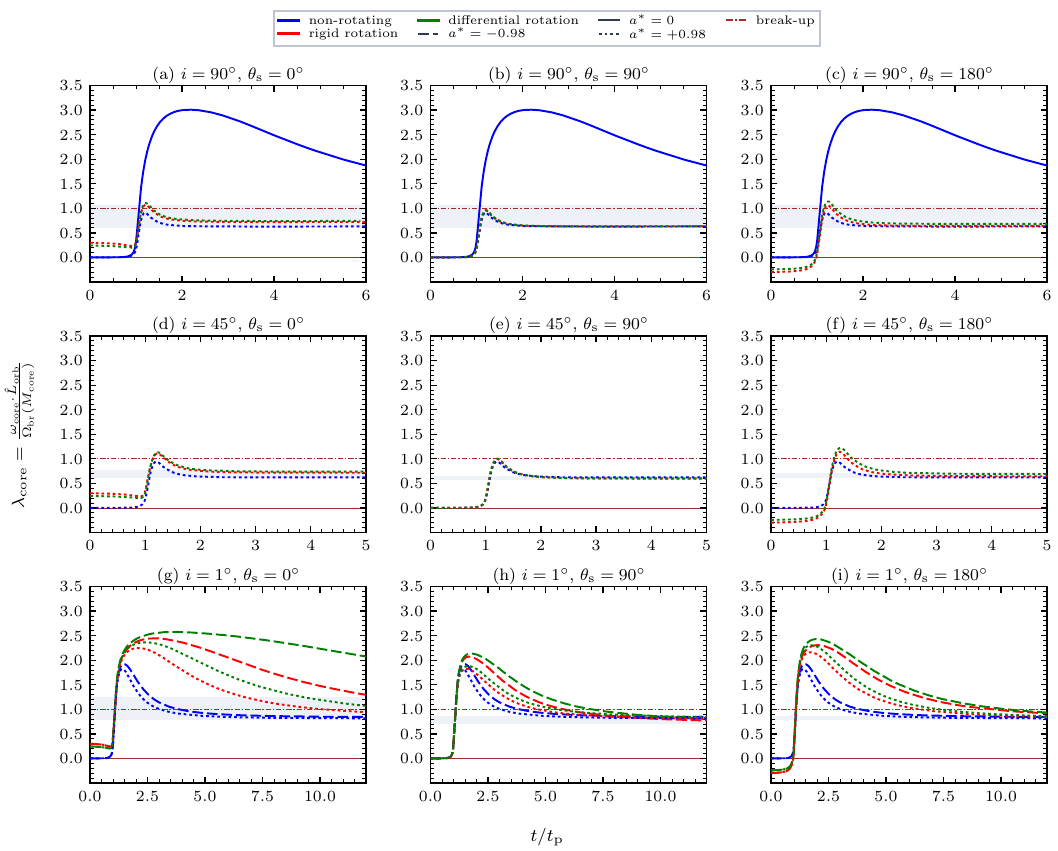}
    \caption{Core spin $\lambda_{\rm core}=\omega_{\rm core}/\Omega_{\rm br}(M_{\rm core})$,
referred to the core after it has regained  equilibrium. Panels, colors and line
styles are as in Figure~\ref{fig:mbound}. The dash-dotted line marks break-up
($\lambda_{\rm core}=1$).  The plots are shown only for the cases that retain more than $5$ percent of the initial stellar mass.}
    \label{fig:spin}
\end{figure}

Figures~\ref{fig:mbound} and \ref{fig:spin} show the bound mass fraction
$M_{\rm bound}/M_\star$ and the core spin. Equation~\eqref{eq:mbound} uses Newtonian quantities: the velocities relative to the core $\mathbf{v}$,
the Newtonian self-gravitational potential and the specific enthalpy. A particle belongs to the core if its specific
energy relative to the core is negative,
\begin{equation}
\label{eq:mbound}
M_{\mathrm{bound}}=\sum_{\epsilon_i<0}m_i,
\qquad
\epsilon_i=\frac{1}{2}\left|\mathbf{v}_i-\mathbf{v}_{\mathrm{core}}\right|^2+\Phi_i+h_i,
\end{equation}
where $\mathbf{v}_{\rm core}$ is the velocity of the core, $\Phi_i$ the self-gravitational
potential, and $h_i=u_i+P_i/\rho_i$ the specific enthalpy. With $h_i$ in place of the
internal energy $u_i$, $\epsilon_i$ is the Bernoulli constant of the gas relative to the
core. In a steady flow, it is conserved along a streamline, so gas with $\epsilon_i>0$ can
escape once its pressure has been converted into kinetic energy \citep{Li2021, Andalman2022}. This criterion is stricter than one based on
$u_i$ and therefore gives smaller cores. The velocities are global coordinate frame velocities. The above definition of energy of the particles bound to the core is Newtonian. So it does not hold good at the pericentre. Near pericentre, they differ from velocities measured by a local Eulerian observer, but we measure the core masses far from the black hole, where the difference between the two sets of velocities is very small. A similar argument applies to the core spin evolution figures. But for qualitative analysis we keep the figures.

In every run the star keeps its full mass until pericentre and loses most of the stripped
mass within $1$--$2\,t_{\rm p}$ after it. We quote core masses at the end of each run. By
then the equatorial and near-polar cores have converged, except the small cores of the
rotating stars on the Schwarzschild orbit, which are still losing mass slowly.

For the non-rotating star the core mass follows the length of the strong-tide phase (Section~\ref{sec:pp}): the star keeps $0.85\,M_\star$ on the prograde equatorial orbit (spends $13.0\,t_\star$ inside $2r_{\rm p}$), $0.36$--$0.40\,M_\star$ on the near-polar orbits ($14.7$--$14.8\,t_\star$), $0.16\,M_\star$ on the Schwarzschild orbit ($15.1\,t_\star$), and nothing on the retrograde equatorial orbit ($22.6\,t_\star$).
For an $n=3/2$
polytrope, the Newtonian threshold for full disruption is $\beta\approx0.9$
\citep{GuillochonRamirezRuiz2013}, so at our $\beta=0.555$ a Newtonian encounter would strip
only a small fraction of the star. The near complete disruptions on the retrograde equatorial and
$45^\circ$ orbits are therefore a strong-field effect. On the near-polar orbit both spin
signs give partial disruptions ($0.40$ and $0.36$). 
For an exactly polar orbit ($L_z=0$) the orbit is neither prograde nor retrograde, and the Kerr symmetry $(\phi,a)\to(-\phi,-a)$ makes the orbits for the two spin signs mirror images of each other, so both would leave the same core. Our orbital plane is inclined by $1^\circ$ to the spin axis, so $L_z$ is small but not zero (see Equation~\eqref{eq:conserved_quantities}): the orbit is slightly prograde for $a^*=+0.98$ and slightly retrograde for $a^*=-0.98$. In Equation~\eqref{eq:rc_rd_parabolic} this enters through the term proportional to $a\sin i$, the only term that changes sign with $a$. The near-polar orbit therefore carries a small part of the difference between the prograde and retrograde equatorial orbits. At $6.3\,r_{\rm g}$ that difference is large (the star spends $13.0$ and $22.6\,t_\star$ inside $2r_{\rm p}$ on the prograde and retrograde equatorial orbits; Section~\ref{sec:pp}), so even this small part separates the two spin signs: the slightly prograde run spends a little less time inside $2r_{\rm p}$ ($14.7$ against $14.8\,t_\star$) and keeps the larger core, and the two cores differ by $0.048\, M_\star$. At $25\,r_{\rm g}$, the prograde and retrograde orbits are already close ($6.3$ and $6.5\,t_\star$), the two near-polar orbits spend the same time inside $2r_{\rm p}$ ($6.4\,t_\star$), and their cores agree to within one per cent (Section~\ref{sec:25rg}). Figure~\ref{fig:mbound} shows that near-polar runs keep more mass than the Schwarzschild run, and on both the star spends slightly less time inside $2r_{\rm p}$ ($14.7$--$14.8$ against $15.1\,t_\star$).

At $6.3\,r_{\rm g}$,  stellar rotation leaves a smaller core compared to the non-rotating star in every
run with surviving core, irrespective of the spin orientation. On the prograde equatorial and $45^\circ$ orbits, the rotating cores are only slightly smaller than the NR core. On the
near-polar orbit they are smaller by up to three quarters, and on the Schwarzschild orbit,
where the NR core is already small, every rotating core is smaller by at least half.
 Rotation therefore matters most when the encounter already leaves the star close to full
disruption. These are also the orbits that keep the star longest inside $2r_{\rm p}$ among those that leave a core.
In Newtonian simulations of rotating stars, aligned rotation strips more mass and anti-aligned rotation protects the star, giving the increasing  order in bound core mass
$0^\circ<90^\circ\lesssim{\rm NR}<180^\circ$ \citep{Golightly2019, Kagaya2019,
SacchiLodato2019, GarainSarkar2025, Mahapatra2026a}. Our runs at $25\,r_{\rm g}$ follow the same order
(Section~\ref{sec:25rg_mass}). 
At $6.3\,r_{\rm g}$
the order holds for aligned and in-plane rotation but not for anti-aligned rotation. The tidal torque transfers orbital angular momentum to the spin of the star, and its effect depends on how the star rotates before the encounter. We describe each case below, using
the core spin $\lambda_{\rm core}$ shown in Figure~\ref{fig:spin} and defined later in this section.

\paragraph{Non-rotating star (NR)} In NR case, the star starts with no spin, and the tidal torque spins it up in the same direction as its orbital motion. Its core mass is set by the duration of the strong tide phase, as described above.

\paragraph{Aligned rotation ($\theta_{\rm s}=0^\circ$)}
In this configuration, the star already rotates in the same direction as its orbital motion, so the tidal torque adds to
its spin. This rotation adds to the velocity that the tide gives to the envelope, and the centrifugal force weakens the effective self-gravity of the star \citep{Golightly2019,
Kagaya2019, SacchiLodato2019}. Aligned rotation therefore gives the smallest core at fixed $a^*$ and $i$ wherever a core survives. On the grazing prograde orbits (the equatorial and $45^\circ$ prograde orbits, which leave very large cores) the orientation changes the core mass by at most about 10 per cent, but even there the aligned star loses more mass than the NR star (Figure~\ref{fig:mbound}).
\paragraph{In-plane rotation ($\theta_{\rm s}=90^\circ$)}
The stellar spin lies in the orbital plane, so the star starts, like the NR star, with no spin about the orbital axis, yet its core is still smaller than the NR core. On the near-polar orbit it is larger than the aligned and anti-aligned cores. As the tide stretches the star, the in-plane rotation adds velocity to one side of the stretched envelope and opposes it on the other along the direction to black hole. The side that gains velocity is prone to getting stripped more easily, so the star loses
more mass overall than the NR star.

\paragraph{Anti-aligned rotation ($\theta_{\rm s}=180^\circ$)}
In anti-aligned case, tidal torque first slows  spin of the star
\citep{SacchiLodato2019, GarainSarkar2025}. At both pericentres the torque reverses the spin shortly before pericentre, before the star starts to lose mass: $\lambda_{\rm core}$ of the
anti-aligned runs passes through zero just before pericentre (Figure~\ref{fig:spin}). Because the reversal happens at both pericentres, the torque cannot by itself explain why
anti-aligned rotation protects the star at $25\,r_{\rm g}$ but not at $6.3\,r_{\rm g}$; nor
can centrifugal support alone. What separates the two is how long the strong tide lasts
after pericentre: $3.2\,t_\star$ at $25\,r_{\rm g}$ against $6.5$--$7.6\,t_\star$ at
$6.3\,r_{\rm g}$ (Section~\ref{sec:pp}). Once reversed, the core rotates with the orbit, so its rotation, and the centrifugal
support it provides, now aids the tidal stripping.
At $25\,r_{\rm g}$ the reversed core reaches
break-up after the others and spends less time above break-up while the tide is still strong, so it keeps the largest core (Section~\ref{sec:25rg_spin}). At
$6.3\,r_{\rm g}$ the strong tide lasts much longer after pericentre, and apsidal precession turns the tidal field through $230^\circ$--$280^\circ$ inside $2r_{\rm p}$
(Section~\ref{sec:pp}). The reversed core therefore spends most of the strong tide rotating with the orbit. The longer strong tide that follows spins every core forward and strips it more. The anti-aligned rotating stars on the prograde equatorial orbit keeps almost as much mass as the non-rotating star but mass striping is noticable. On the near-polar orbit the contrast with the $25\,r_{\rm g}$ cases for rotating stars against NR star is significant. Unlike at $25\,r_{\rm g}$, anti-aligned rotation offers no protection at this pericentre.

\paragraph{Difference between RR and DR stellar models}
The DR star leaves a slightly smaller core than the RR star in every near-polar run and on the prograde equatorial and $45^\circ$ orbit with $\theta_{\rm s}=180^\circ$; elsewhere the two agree closely. The difference lies in the angular velocity profile of the DR star. In the DR star the inner region rotates much faster than the envelope, so the centrifugal force in the rapidly rotating inner layers makes the DR star less centrally concentrated, which lowers the binding of its slowly rotating outer layers. These outer layers are therefore stripped more easily than in the RR star, and
the core that remains is made of the rapidly rotating inner region, consistent with the slightly faster spin of the DR cores described below.

The core spin has the same pattern in every run (Figure~\ref{fig:spin}). We define a spin diagnostic term, $\lambda_{\rm core}=\omega_{\rm core}/\Omega_{\rm br}(M_{\rm core})=L_s/(I_0M_{\rm core}^{4/3})$
in units of $M_\star$ and $R_\star$,where $L_s$ is the spin angular momentum of the bound particles about the centre of mass of
the core, $I_0$ is the dimensionless moment of inertia of the relaxed star in units of
$M_\star R_\star^2$ (0.205 for the non-rotating star, as for an $n=3/2$ polytrope, and
0.222 and 0.221 for the RR and DR stars, which rotation flattens), and the factor
$M_{\rm core}^{4/3}$ rescales it to an $n=3/2$ polytrope with the mass of the core. The
core is tidally deformed and does not rotate rigidly, so $\omega_{\rm core}$ is not its
instantaneous rotation rate but its mean angular velocity, weighted by moment of inertia,
once it has settled into equilibrium with its present angular momentum. $\lambda_{\rm core}$
compares this mean rate with the break-up rate of the settled core: a value above unity
means that the core cannot settle with its present spin and must still shed mass or
angular momentum. The tidal torque spins the core up at pericentre, and the core then spins
down as its fast outer layers are stripped. The depth of the encounter sets the amplitude.
Grazing prograde encounters leave the core below
break-up, while deep partial disruptions (the near-polar and Schwarzschild orbits) drive
$\lambda_{\rm core}$ to two or three times break-up before it relaxes towards break-up. The deep cores cross break-up near the end of the strong-tide phase and reach their peak spin after it, so the slow mass loss that follows in Figure~\ref{fig:mbound} is the over-spun core shedding its outer layers. On the grazing orbits the spin peaks close to break-up near the end of the strong-tide phase, and the core mass hardly changes afterwards. For
the NR star, the peak spin rises steadily from $0.9$ to $3.0$ as the bound mass falls from
$0.85$ to $0.16\,M_\star$.

The initial rotation leaves only a small imprint on the core spin. Rotating cores end
slightly faster than NR cores in the grazing encounters with $\theta_{\rm s}=0^\circ$, and
reach higher peaks on the near-polar orbit, and DR cores spin slightly faster than RR cores.
The final spin of a surviving core is therefore set by the encounter rather than by the
star.

\subsection{Debris energy distribution}
\label{sec:dmde}

\begin{figure}[tbp]
    \includegraphics[width=1.0\textwidth]{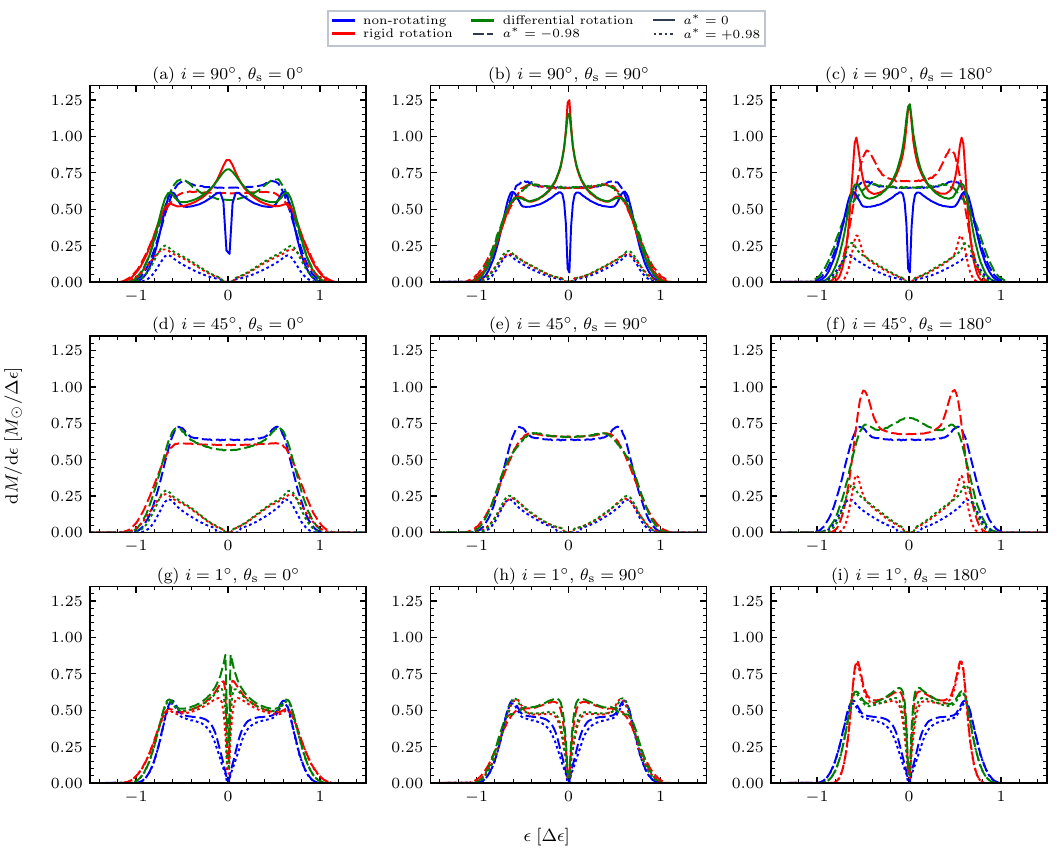}
    \caption{Energy distribution of the stripped debris, $dM/d\epsilon$, in units of
    $\Delta\epsilon=GM_{\rm BH}R_\star/r_{\rm t}^2$. Bound cores above $0.05\,M_\star$ are excluded; the smaller bound remnants on the Schwarzschild and $45^\circ$ retrograde orbits are kept, and on the Schwarzschild orbit they give the central peaks in panels (a)--(c).
    Panel colours and line styles are as in Figure~\ref{fig:mbound}.
    The snapshots are taken sufficiently long after pericentre passage, by which time the energy distributions have frozen, i.e., become time-independent.}
    \label{fig:dmde}
\end{figure}

Figure~\ref{fig:dmde} shows the distribution of the stripped debris over specific orbital
energy, $dM/d\epsilon$, in units of $\Delta\epsilon=GM_{\rm BH}R_\star/r_{\rm t}^2$. In
all runs, the distribution is close to symmetric about $\epsilon=0$, with $49$ and $51$ per cent of the debris is bound to the black hole, consistent with the small
asymmetry between the bound and unbound tails found by \citet{Banerjee2023}. Neither the black hole spin, nor the inclination, nor the stellar spin shifts the debris measurably towards bound or unbound
orbits.

We define the width of the distribution as the energy $|\epsilon|$ at which $dM/d\epsilon$
falls to 5 per cent of its maximum. For the non-rotating star it is
$0.88$--$0.95\,\Delta\epsilon$ in every run, from the prograde equatorial encounter, which
leaves most of the star bound, to the retrograde one, which destroys it. This is consistent with the energy spread being set near the
tidal radius rather than at pericentre \citep{StoneSariLoeb2013}. The black hole spin
therefore sets how much mass is stripped, but hardly changes the range of energies over
which the debris is spread. 

The surviving core leaves an imprint near $\epsilon=0$, where the debris least bound to the black hole lies.
In the full disruptions, $dM/d\epsilon$ is flat across $\epsilon=0$. In partial disruptions, the core removes the
debris near $\epsilon=0$ and leaves a depleted region whose width grows with the core
mass. The region is widest for the massive prograde cores ($M_{\rm bound}/M_\star>0.7$),
where $dM/d\epsilon$ falls steadily to zero from its peak, and narrow for the near-polar
cores. The small cores of the rotating stars on the near-polar orbit leave a narrow notch
 bordered by excess mass, which is likely gas shed
by the rapidly spinning core after pericentre. Because the least bound debris returns last,
this region controls the late-time fallback rate \citep{CoughlinNixon2019}.

Stellar rotation changes the width of the distribution. Relative to the non-rotating star,
aligned RR rotation widens it and anti-aligned RR rotation narrows it, each by
$11$--$21$ per cent, on every orbit and for every black hole spin, while in-plane rotation
changes it by less than 10 per cent. Rotation adds to the orbital velocity on one side of
the star and subtracts from it on the other, which increases the energy spread for aligned
rotation and decreases it for anti-aligned rotation \citep{Golightly2019}. Multiplying the
surface rotation speed by the orbital speed at pericentre gives an expected shift
$\delta\epsilon/\Delta\epsilon\approx\lambda_{\rm rot}(2\beta)^{1/2}\approx0.3$ for RR,
comparable to the measured $0.1$--$0.2\,\Delta\epsilon$. The narrower anti-aligned
distributions also develop sharper peaks. The DR width differs from the NR width by at most
10 per cent, because the high-$|\epsilon|$ tails come from the envelope, which rotates slowly
in the DR model ($\Omega\simeq0.07\,\Omega_{\rm br}$ at the surface, against
$0.3\,\Omega_{\rm br}$ for RR).

\subsection{Fallback rate and its late-time slope}
\label{sec:fallback}

\begin{figure}[tbp]
    \includegraphics[width=1.0\textwidth]{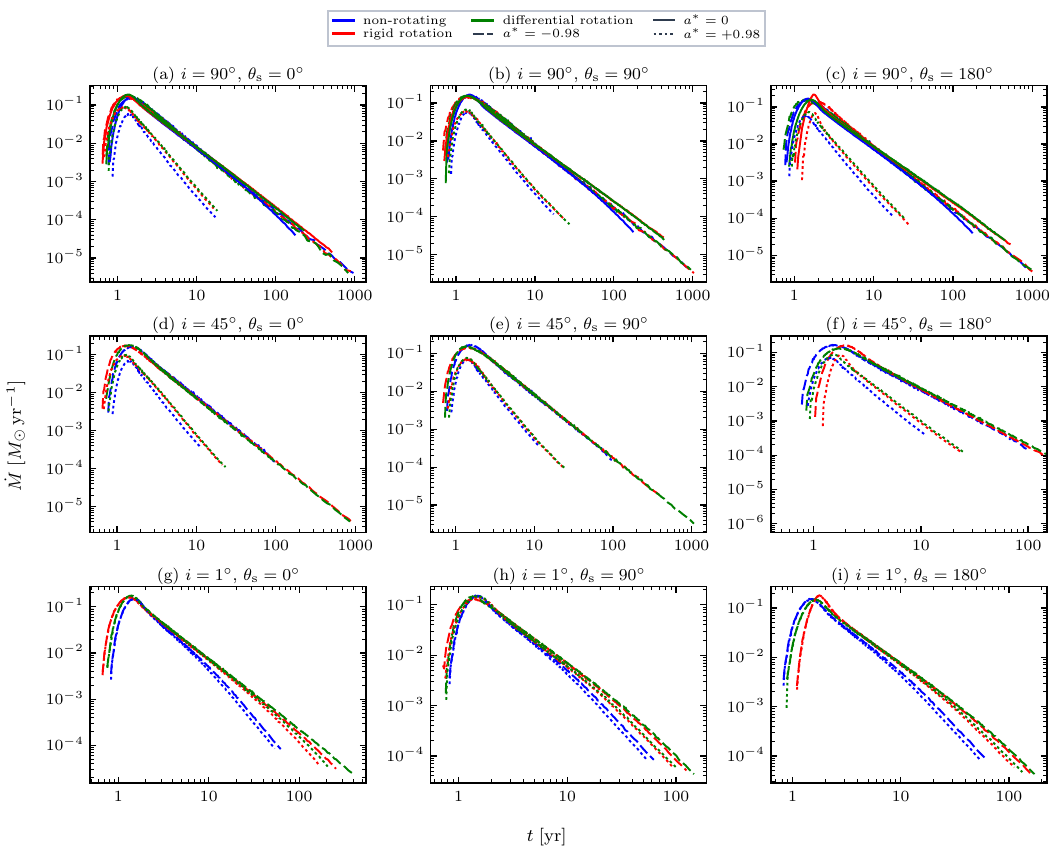}
    \caption{Fallback rate $\dot M$ in the frozen-in approximation, with time measured from
pericentre passage. Panels, colours and line styles are as in Figure~\ref{fig:mbound}.}
    \label{fig:fallback}
\end{figure}

\begin{figure}[tbp]
    \includegraphics[width=1.0\textwidth]{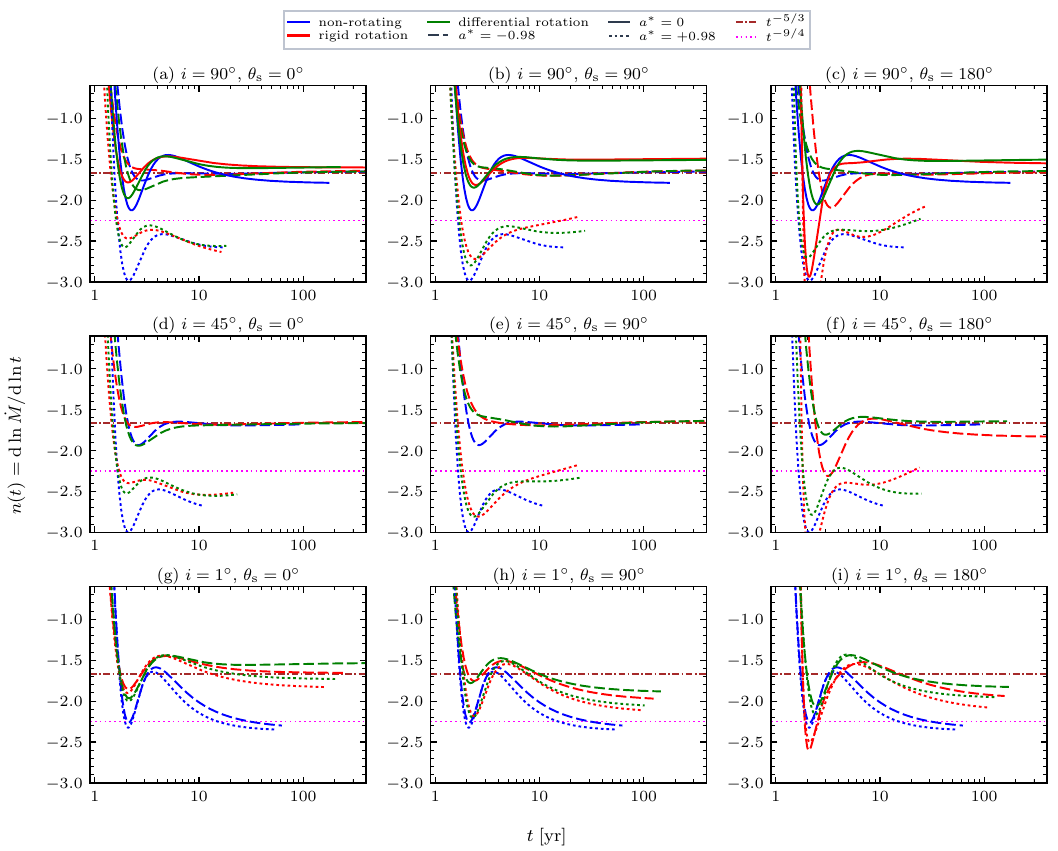}
    \caption{Slope of the fallback rate, $n=d\ln\dot M/d\ln t$. Panels, colours and line styles
are as in Figure~\ref{fig:mbound}. The dash-dotted and dotted horizontal lines mark
$n=-5/3$ and $n=-9/4$.}
    \label{fig:slope}
\end{figure}

Figures~\ref{fig:fallback} and \ref{fig:slope} show the fallback rate $\dot M$ and its
logarithmic slope $n\equiv d\ln\dot M/d\ln t$. We compute $\dot M$ in the frozen-in approximation from a late snapshot of each simulation, taken after the debris energies have frozen: $dM/d\epsilon$ changes little over the late snapshots of each run. For every debris particle that is not bound to the surviving core and is bound to the black hole, we compute its return time, measured from pericentre passage, as the Keplerian period
\begin{equation}
t_{\rm ret}=2\pi\left(\frac{a^3}{GM_{\rm BH}}\right)^{1/2},\qquad a=\frac{GM_{\rm BH}}{2|\epsilon|}.
\end{equation}
For an inclined Kerr orbit the coordinate-time period depends on the polar as well as the radial motion, so it would require integrating the $\theta$ equation for every bound debris particle, which is computationally expensive for the large number of particles that contribute to the fallback.
Relativistic periods give almost the same fallback rate, because the debris spends nearly all of its orbit far from the black hole. The fallback rate is the mass returning per unit time, obtained by binning $t_{\rm ret}$ in linear bins of width $t_{\rm peak}/80$ from the first return until bin would contain less than 300 particles, and in bins of 300 particles thereafter. Each curve ends when 99 per cent of the bound debris has returned.  We fit binned rate with modified Padé approximation of \citet{Nixon2021} ($N_{max} = 5$) in log space, $n(t)$ in Figure \ref{fig:slope} is the derivative of this fit and we quote its asymptotic index $n_{\infty}$ as late time slope.

In all runs the fallback rate peaks at $t_{\rm peak}\simeq1.2$--$2.0$\,yr with
$\dot M_{\rm peak}\simeq0.06$--$0.22\,M_\odot\,{\rm yr}^{-1}$. For a radiative efficiency of
$0.1$ the Eddington rate of a $5\times10^7\,M_\odot$ black hole is
$\simeq1.1\,M_\odot\,{\rm yr}^{-1}$, so the peaks lie $5$--$18$ times below it. The peak time
corresponds to the maxima at the bound part of $dM/d\epsilon$, and the peak rate follows the stripped mass,
so the massive prograde cores give the lowest peaks.

Full disruptions approach the canonical $t^{-5/3}$ law \citep{Rees, Phinney1989}, but only
after a transient that depends on the stellar spin. Shortly after the peak the slope falls
below $-5/3$ and then recovers. The dip is deepest for anti-aligned RR rotation, smaller for
the non-rotating star and for aligned DR rotation, and almost absent for aligned RR and
in-plane rotation. In the frozen-in calculation, the dip follows from the shape of
$dM/d\epsilon$: applied locally, Equation~\eqref{eq:nk} gives a slope below $-5/3$ where
$dM/d\epsilon$ falls towards $\epsilon=0$ just inside its peaks, so the sharper peaks of the
anti-aligned distributions (Section~\ref{sec:dmde}) give the deepest dip. On the retrograde
equatorial orbit, every run then settles onto $t^{-5/3}$, within a few years to a few decades
of the disruption, and stays within $0.03$ of it to the end of our curves, with no extended
phase flatter than $t^{-5/3}$. The same holds on the $45^\circ$ retrograde orbit, which
leaves a remnant of $\lesssim0.03\, M_\star$, except for the two anti-aligned runs. After its
dip the anti-aligned DR run bounces back to a decline flatter than $t^{-5/3}$ and needs about
a decade to settle, much as \citet{Golightly2019} found for their full disruptions within years. The
anti-aligned RR run also bounces back, but its small remnant then steepens the
tail steadily, so even a very small core can prevent a $t^{-5/3}$ decline.
\citet{Golightly2019} also found that the dip is deeper for retrograde stellar spin and
attributed it to self-gravitating clumps in the debris stream; our frozen-in calculation does
not follow the stream, so in our runs the dip reflects the energy distribution alone. At
late times ($t\gtrsim100$\,yr) all full disruptions except the anti-aligned RR run on the $45^\circ$ retrograde orbit decline slightly more slowly than
$t^{-5/3}$, by at most $0.03$ in $n$, but this also holds for the non-rotating star, so we
cannot attribute it to stellar rotation.

Partial disruptions have no universal late-time slope. Since the late-returning debris
follows the Keplerian scaling $t\propto|\epsilon|^{-3/2}$, a distribution
$dM/d\epsilon\propto|\epsilon|^{k}$ near $\epsilon=0$ gives a late-time slope
\begin{equation}
n\simeq-\frac{5}{3}-\frac{2k}{3}.
\label{eq:nk}
\end{equation}
A flat distribution ($k=0$) gives $t^{-5/3}$. A distribution that falls linearly to zero
($k=1$), as near a core that has removed the least bound debris, gives $t^{-7/3}$. Excess
mass near $\epsilon=0$ ($k<0$) gives a decline shallower than $t^{-5/3}$. Our partial
disruptions span this whole range, and for cores of $M_{\rm bound}/M_\star\le0.40$ the
late-time slope steepens with core mass. For the near-polar NR cores $dM/d\epsilon$ has
$k\simeq1.0$ near $\epsilon=0$, so Equation~\eqref{eq:nk} gives $n\simeq-2.33$, in agreement
with the measured $n\simeq-2.3$. Small, rapidly spinning cores keep shedding gas onto
marginally bound orbits ($k<0$), so their tails are flatter than $t^{-5/3}$. The massive
prograde cores keep almost all the debris with small $|\epsilon|$, so their fallback ends
within a few decades, whereas the smaller cores keep feeding the black hole for hundreds of
years. On the near-polar orbit the $a^*=-0.98$ run leaves the smaller core and has the
shallower tail for every stellar model. \citet{CoughlinNixon2019} predict $t^{-9/4}$ for
partial disruptions, a slope that arises from the gravity of the core on the returning
debris.

Rotation changes the fallback rate in two ways. It shifts the peak in the same sense as it
changes the width of $dM/d\epsilon$ (Section~\ref{sec:dmde}): aligned rotation advances the
peak , anti-aligned rotation delays it, and in-plane rotation leaves it unchanged, as
found by \citet{Golightly2019}. In partial disruptions, rotation also strips more mass, which
raises the peak on the prograde orbits and flattens the late-time decline on the near-polar and Schwarzschild orbits.
The slowly rotating envelope of the DR
star weakens the first effect, and its smaller cores strengthen the second: anti-aligned DR
delays the peak less than RR, and on the near-polar orbit DR gives a shallower tail than RR
in all six runs.

The spin orientation changes the late-time slope through the shape of $dM/d\epsilon$ near
$\epsilon=0$. On the near-polar orbit it orders the slope as it orders the core mass: aligned
rotation, which leaves the smallest core, gives the shallowest decline, in-plane rotation
gives the steepest, and anti-aligned rotation lies in between. On the prograde equatorial
and $45^\circ$ orbits, where all cores are massive, aligned RR instead declines most
steeply of the RR runs, and the DR runs follow the same order on the equatorial orbit but not on the $45^\circ$ orbit.

The fallback rate traces the luminosity only while the returning debris is accreted faster
than it arrives, that is, while the viscous time of the disc it forms is short compared with
the age of the event \citep{Rees, LodatoRossi2011, ShenMatzner2014, Golightly2019}. At our
pericentre, strong apsidal precession makes the stream collide with itself close to the black
hole, so the debris should circularise promptly at about $2r_{\rm p}$ \citep{Hayasaki2013,
Dai2015, Bonnerot2016}, although Lense-Thirring precession can delay this on inclined orbits
\citep{StoneLoeb2012, Hayasaki2016}. For a viscosity parameter $\alpha=0.1$
\citep{ShakuraSunyaev1973, Pringle1981}, the viscous time at this radius is about $0.3$\,yr
for a disc aspect ratio $H/R=0.1$, shorter than $t_{\rm peak}$. Near the peak the disc is
supported by radiation pressure resulting in a thick disc. As the supply declines, the disc becomes
thinner and its viscous time grows, to about $9$\,yr for $H/R=0.02$ and $35$\,yr for
$H/R=0.01$. After about a decade, the viscous time therefore exceeds the age of the event: the returning gas accumulates in the disc, and the light curve follows the viscous spreading of
the disc rather than the fallback rate \citep{Cannizzo1990, ShenMatzner2014}, as in the
late-time plateaus of observed TDEs \citep{vanVelzen2019}. We therefore take the first
$10$\,yr, by which time at least two thirds of the bound debris has returned, as the window in
which our fallback rates can be compared with light curves. Within this window the full disruptions already decline close to $t^{-5/3}$, except for the anti-aligned runs on the $45^\circ$ retrograde orbit, and the partial disruptions already differ: the fallback rate behind the massive prograde cores falls off much more steeply, while on the near-polar and Schwarzschild orbits the late-time slope steepens with core mass.
 Beyond this window, our curves describe the mass supply to the black
hole rather than the light curve, and they are increasingly affected by low bin resolution.

\section{Comparison with weak-field disruption results at $r_{\mathrm p}\simeq25\,r_{\rm g}$}
\label{sec:25rg}

In this section, we present the TDE results of
the equatorial ($i=90^\circ$) and near-polar ($i=1^\circ$) simulations with
$M_{\rm BH}=4\times10^{6}\,M_\odot$ and $\beta=0.75$. This configuration places the pericentre at
$r_{\rm p}\simeq25\,r_{\rm g}$. The simulations for $45^\circ$ orbits for this weak encounter are not shown, as these sit in between near-polar and equatorial orbits and do not show any different trends. In the weak-field limit, the simulation with $\beta =0.555$ for BH mass of the order $\leq 1.5\times 10^{7}\,M_{\odot}$ mostly results in a grazing encounter or no disruption at all. The stellar models and spin orientations are those of
Section~\ref{sec:results}. This set closely matches the encounters of
\citet{Mahapatra2026a} ($r_{\rm p}\simeq25\,r_{\rm g}$, $\beta=0.7$, $a^*=\pm0.98$). All
comparisons below are made at fixed orbit and fixed black hole spin.

\begin{figure}[!tbp]
    \includegraphics[width=\textwidth]{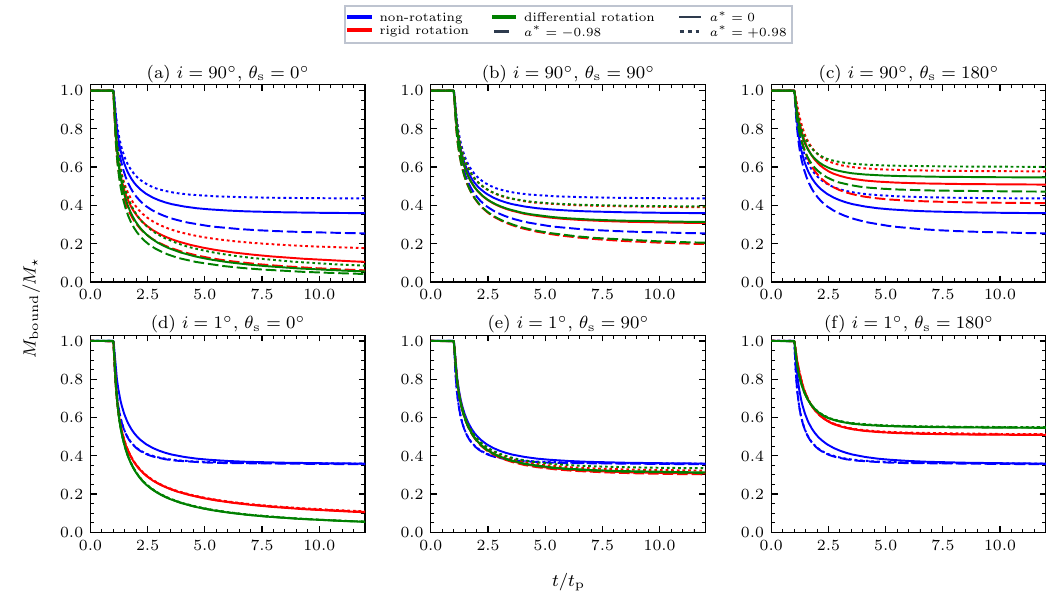}
    \caption{Bound mass fraction $M_{\rm bound}/M_\star$ at $r_{\rm p}\simeq25\,r_{\rm g}$.
    The top row shows the equatorial orbit ($i=90^\circ$) and the bottom row the near-polar
    orbit ($i=1^\circ$). Columns show $\theta_{\rm s}=0^\circ$, $90^\circ$ and $180^\circ$.
    Colours and line styles are as in Figure~\ref{fig:mbound}.}
    \label{fig:mbound_25rg}
\end{figure}

\subsection{Bound core mass}
\label{sec:25rg_mass}

The top and bottom panels of Figure~\ref{fig:mbound_25rg}  show the bound core mass in equatorial and off-equatorial orbits, respectively, at a pericenter distance of $25\,r_{\rm g}$. For
the non-rotating star the black hole spin matters on the equatorial orbit, where the core
grows from $a^*=-0.98$ to $+0.98$, but hardly on the near-polar orbit, whose cores lie
between the two equatorial ones, as in \citet{Mahapatra2026a}. The stellar spin orientation
has a much larger effect and gives the same  increasing order of core mass post disruption on both orbits and for every black hole
spin, $0^\circ<90^\circ\lesssim{\rm NR}<180^\circ$. Aligned rotation removes most of the rotating stars'
core, while anti-aligned rotation protects the spinning stars and keeps the largest core in every
configuration. For both equatorial and polar orbits the effect of stellar rotation in bound core is significant compared to NR cases. This result also agrees with the order found in Newtonian SPH
simulations of rotating stars \citep{Golightly2019, Kagaya2019, SacchiLodato2019,
GarainSarkar2025} and by \citet{Mahapatra2026a}. At $6.3\,r_{\rm g}$ the stabilizing effect of anti-aligned rotation is lost:
Every rotating core, including the anti-aligned one, is smaller than the NR core
(Section~\ref{sec:core}).

\subsection{Core spin and anti-aligned protection}
\label{sec:25rg_spin}

\begin{figure}
    \includegraphics[width=\textwidth]{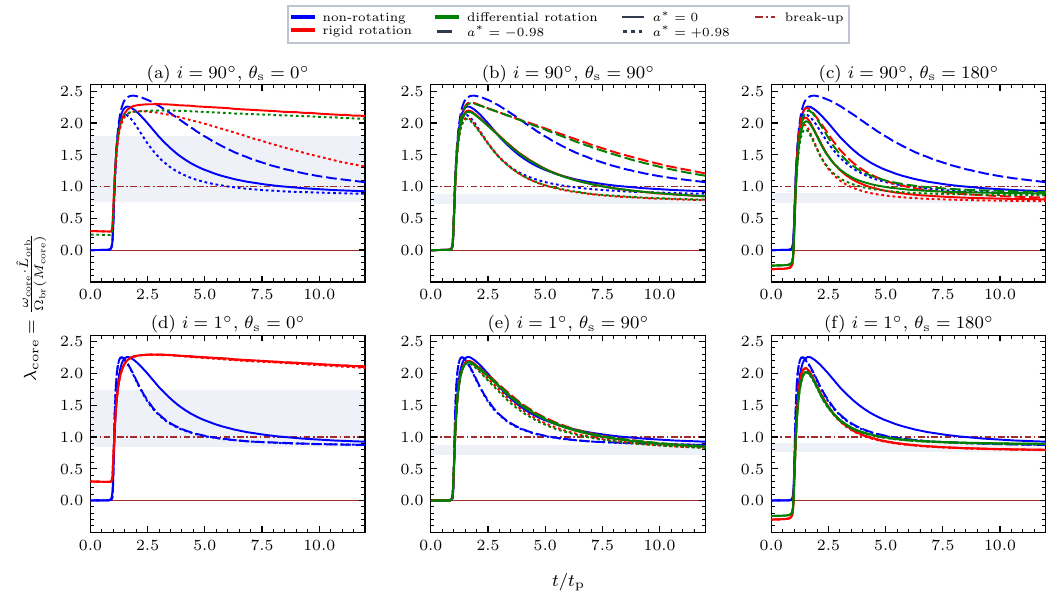}
    \caption{Core spin about the orbital axis at $r_{\rm p}\simeq25\,r_{\rm g}$. Panels are as
    in Figure~\ref{fig:mbound_25rg}, and colours and line styles as in
    Figure~\ref{fig:mbound}. The dash-dotted line marks break-up ($\lambda_{\rm core}=1$) and
    the shaded band spans the final values. The plots are shown only for the cases that retain
    more than $2$ percent of the initial stellar mass.}
    \label{fig:spin_25rg}
\end{figure}

Figure~\ref{fig:spin_25rg} shows the evolution of the core spin $\lambda_{\rm core}$ about the orbital axis
at $25\,r_{\rm g}$ on the equatorial (top) and near-polar (bottom) orbits. The tidal torque spins the aligned core up more than the NR core and first slows the anti-aligned spin
\citep{SacchiLodato2019, GarainSarkar2025}, reversing it shortly before pericentre, before
the star starts to lose mass. The anti-aligned core therefore reaches break-up after the
others. The strong tide lasts only about $2\,t_\star$ after the cores pass break-up, so the
anti-aligned core spends the least time above break-up while the tide is still strong. Most
of the mass is lost afterwards, as the spun up cores shed their outer layers, and this late mass loss follows the same order: $0^\circ<90^\circ\lesssim{\rm NR}<180^\circ$.
The same reversal occurs at $6.3\,r_{\rm g}$, but there the strong tide lasts $6.5$--$7.6\,t_\star$ after pericentre and anti-aligned rotation no longer protects the star (Section~\ref{sec:core}).

\subsection{Debris morphology}
\label{sec:25rg_morph}
Our simulations capture several morphological features of the tidal debris streams, some of which have been reported in previous studies \citep{GaftonRosswog2019}.

\begin{figure}[tbp]
    \includegraphics[width=\textwidth]{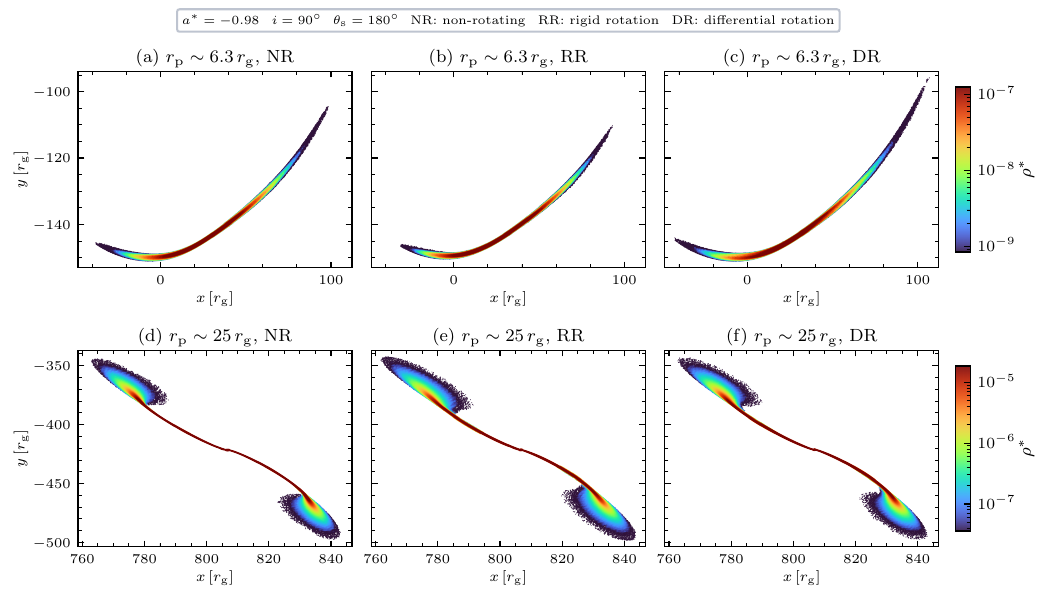}
    \caption{Debris on the equatorial orbit ($i=90^\circ$, $a^*=-0.98$, $\theta_{\rm s}=180^\circ$), projected on the $x$--$y$ plane, for $r_{\rm p}\simeq6.3\,r_{\rm g}$ (top) and $r_{\rm p}\simeq25\,r_{\rm g}$ (bottom). Columns show the NR, RR and DR models. Colour bar gives the conservative density $\rho^*$ on a logarithmic scale. The diffuse low-density gas around a surviving core is not shown. All snapshots are taken at $t=200\,t_{\star}.$}
    \label{fig:morph_eq}
\end{figure}

Figure~\ref{fig:morph_eq} shows the debris of retrograde equatorial orbit for NR, RR and DR stellar models with stellar spin anti-aligned to orbital angular momentum. At $6.3\,r_{\rm g}$ every stellar model is fully disrupted into a single curved stream with no bound core. At $25\,r_{\rm g}$ bound core survives in all three models. Distinct tidal arms extend into two debris tails, each of which broadens into a diffuse, low-density tidal lobe. All stellar models produce very similar morphologies on both equatorial and near-polar orbit.

\begin{figure}[htbp]
    \includegraphics[width=1.0\textwidth]{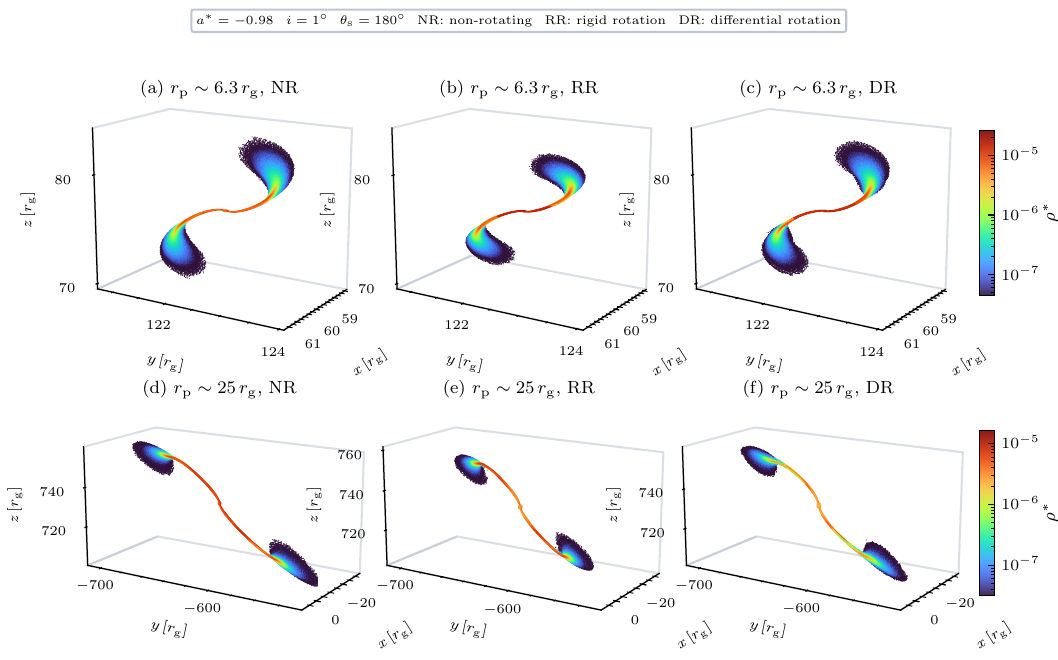}
    \caption{Three dimensional view of the debris on the near-polar orbit ($i=1^\circ$, $a^*=-0.98$, $\theta_{\rm s}=180^\circ$) for $r_{\rm p}\simeq6.3\,r_{\rm g}$ (top) and $r_{\rm p}\simeq25\,r_{\rm g}$ (bottom). Columns and colours are as in Figure~\ref{fig:morph_eq}. The axes are scaled separately so that the debris fills each panel, so the shapes are stretched. All snapshots are taken at $t=200\,t_{\star}$.}
    \label{fig:morph_offeq}
\end{figure}

Figure~\ref{fig:morph_offeq} shows the debris on the near-polar orbit. At both pericentres the stream is $\mathrm{S}$-shaped with a bound core at its centre, but difference can be observed in tidal lobes. At $25\,r_{\rm g}$, the lobes remain relatively compact, whereas at $6.3\,r_{\rm g}$, they are more extended and curl back towards the stream in opposite directions. This difference is consistent with the stronger Lense-Thirring precession of the near-polar orbit at $6.3\,r_{\rm g}$ (Section~\ref{sec:pp}), Since the precession depends on the time spent near the black hole, gas stripped at different stages of the encounter can experience different orbital-plane tilts, potentially contributing to the observed curvature of the lobes. However, the two sets of runs also differ in $\beta$, apsidal precession and the duration of the strong tide. These differences prevent us from isolating the contribution of Lense-Thirring precession or attributing the curled lobe morphology to it alone.

\section{Discussions and Conclusions}
\label{sec:conclusion}
We have performed general-relativistic SPH simulations of the tidal disruption of a
solar-type $\gamma=5/3$ polytrope by a Kerr black hole with $a^*=\pm0.98$, using the
Schwarzschild case ($a^*=0$) as a reference. In the strong-field runs the black hole mass, $M_{\mathrm{BH}}= 5\times10^{7}\,M_\odot$ and the pericentre is $r_{\rm p}\simeq6.3\,r_{\rm g}$. As a comparison with weak-field simulations, we repeated the encounters with
$r_{\rm p}=25\,r_{\rm g}$ for $M_{\mathrm{BH}}= 4\times10^{6}\,M_\odot$. We considered equatorial
($i=90^\circ$), inclined ($i=45^\circ$) and near-polar ($i=1^\circ$) orbits for the close encounter ($r_{\rm p}\simeq6.3\,r_{\rm g}$. For weak encounter ($r_{\rm p}\simeq25\,r_{\rm g}$, only equatorial and near-polar orbits are shown. For each orbit
we compared a non-rotating (NR) star, a rigidly rotating (RR, $\lambda_{\rm rot}=0.3$) star
and a differentially rotating (DR, Hachisu law with $\lambda_{\rm rot}=1.8$, $A=0.2$) star,
with the stellar spin inclined at $\theta_{\rm s}=0^\circ$, $90^\circ$ and $180^\circ$ to
the orbital angular momentum. Unless stated otherwise, the values quoted below refer to
$r_{\rm p}\simeq6.3\,r_{\rm g}$ and are measured at the end of each run.
A central finding of our study is that the influence of stellar rotation on tidal disruption changes substantially between the weak and strong-field regimes. At $r_{\rm p}\simeq25\,r_{\rm g}$, the surviving core mass follows the ordering $0^\circ<90^\circ<{\rm NR}<180^\circ$ for all orbital inclinations and all black hole spin configurations considered. This is consistent with the Newtonian picture in which anti-aligned stellar rotation provides additional resistance to tidal deformation, allowing the star to retain much larger bound mass than its non-rotating counterpart. At $r_{\rm p}\simeq6.3\,r_{\rm g}$, this behaviour is reversed as all rotating configurations retain less mass than the non-rotating star. The reversal is closely connected to the post-pericentre evolution of the stellar spin as explained in Section~\ref{sec:25rg_spin}. The interaction between black hole spin and stellar spin also plays a decisive role in determining the survival of the stellar core in the strong-field regime. For our simulation regime, equatorial and $45^\circ$ encounters, retrograde configurations result in complete or nearly complete disruption, where as prograde orbits irrespective of the rotation profile are grazing encounters. The dependence on black hole spin is less pronounced for near-polar encounters, where all spin orientations produce partial disruptions, although their surviving core masses differ by approximately $0.05\,M_\star$. In comparison, the corresponding weak-field encounters at $r_{\rm p}\simeq25\,r_{\rm g}$ retain core masses that agree to within one per cent. These results highlight the importance of black hole spin and orbital inclination in determining the disruption outcome when the pericentre approaches the strong-field region.
We also find a modest but noticeable difference between the two rotation profiles. In case of  the near-polar orbit, the differentially rotating star loses slightly more mass than its rigidly rotating counterpart, despite carrying lesser initial angular momentum when they were relaxed. The final core spin depends on how deep the encounter is, not on the initial rotation. Grazing encounters leave cores below break-up ($\lambda_{\rm core}\simeq0.60-$--$0.74$). Deep partial disruptions spin the core up to $\lambda_{\rm core}\approx2$--$3$, after which subsequent shedding of mass and angular momentum brings the remnant towards rotational break-up.  
The effects of black hole and stellar rotation also differ in their influence on the debris energy distribution and the resulting fallback history. While the black hole spin substantially changes the amount of stripped mass, it has a comparatively weak influence on the energy spread of the debris. Stellar rotation, on the other hand, produces a clear systematic effect. Aligned rotation broadens the debris energy distribution and advances the fallback peak, whereas anti-aligned rotation narrows the distribution and delays the peak. For the encounters considered here, the fallback rate reaches its maximum approximately $t\simeq1.2$--$2.0$\,yr after disruption, with $\dot M_{\rm peak}\simeq0.06$--$0.22\, M_\odot\,{\rm yr}^{-1}$, at least a factor of five below the Eddington rate. The massive bound cores in prograde encounters produce the lowest peaks ($0.06$--$0.10\,M_\odot\,{\rm yr}^{-1}$), and their fallback ends within $\approx11$--$27$\,yr because the core keeps the least bound debris.
The late-time slope depends strongly on whether a surviving core is present. Complete disruptions approach the canonical $t^{-5/3}$ decline following an initial transient whose duration and shape depend on the stellar spin except for RR with $\theta_{\rm s}=180^\circ$ on the $45^\circ$ retrograde orbit, where the fallback steepens the tail to $n=-1.83$. Partial disruptions, in contrast, do not exhibit a universal late-time power-law index. Their asymptotic behaviour is determined by the structure of the debris energy distribution near $\epsilon=0$, producing slopes ranging from approximately $t^{-1.49}$ to $t^{-2.7}$. The $t^{-9/4}$ law is reached for cores of $\approx0.35$--$0.40\,M_\star$. Small, rapidly spinning cores give late time fallback shallower than $t^{-5/3}$. Our result matches with \citet{Nixon2021} near the partial/full-disruption threshold, where they find late-time slopes between
$-5/3$ and $-9/4$ ($n\simeq-1.7$ to $-1.9$) for marginally surviving or re-formed (`zombie')
cores, and slopes shallower than $-5/3$ ($n\simeq-1.4$ to $-1.6$) just beyond full disruption. Our runs closest to the threshold give $n=-1.49$ to $-1.83$, in the same range;
in their Newtonian simulations these slopes arise from the gravity of the core and of the stream.

TDEs are increasingly used to measure the masses of quiescent supermassive black holes. The late time UV/optical plateau luminosity scales tightly with black hole mass, and the peak optical luminosity and radiated energy follow empirical mass scalings \citep{Mummery2024}. Population models built on these relations reproduce the observed optical, UV and X-ray luminosity functions with black hole mass as the controlling parameter \citep{Mummery2025}, and they are now used to infer black hole masses from the optical flare alone, for both full
and repeating partial disruptions \citep{Mummery2026}.
Our simulations show that, in the strong field regime, black hole mass alone does not set the debris return. Black hole and stellar spin are therefore hidden parameters that are likely to add scatter to the TDE scaling relations, particularly near the Hills mass. This matters more for repeating partial TDEs. A few surviving core in our parabolic orbit simulation is spun up to near or above break-up, so successive flares from the same star should differ in stripped mass, peak time and decay slope even on an unchanged orbit.
Our fallback rates describe only the mass returning to pericentre. The luminosity used in black hole mass inference, however, comes from the disc that forms once this debris circularizes. Natural next steps are to follow the debris through stream collisions and disc formation, as well as the possibility of repeated disruptions of the spun-up cores on subsequent pericenter passages. 
\FloatBarrier
\section*{Acknowledgements}
We acknowledge the support and resources provided by PARAM Sanganak under the National Supercomputing Mission, Government of
India, at the Indian Institute of Technology Kanpur.

\section*{Data Availability}
The data underlying this article will be shared upon reasonable request to the corresponding author.


\FloatBarrier

\appendix
\section{Differential and rigid rotation validation}
\label{appendixA}
\begin{figure}
    \centering
    \begin{minipage}[b]{\textwidth}
        \centering
        \includegraphics[width=\textwidth]{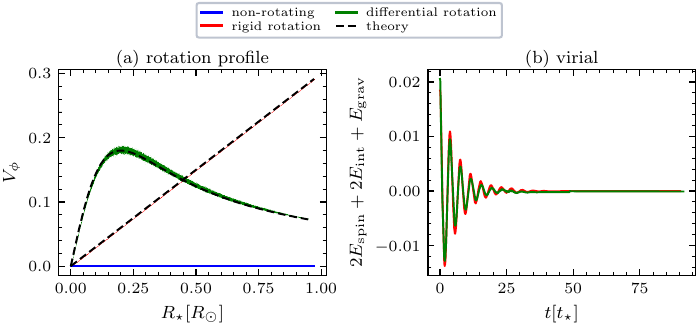}
    \end{minipage}
    
    \caption{Left: azimuthal velocity of the SPH particles against cylindrical radius for
    the relaxed  RR and DR stars created from relaxed NR star, with the target profiles (dashed). Right: virial sum
    $2E_{\rm spin}+2E_{\rm int}+E_{\rm grav}$ during the spin-up and equilibrium test of the
    RR and DR stars.}
\label{fig:star_profiles}
\end{figure}

Figure~\ref{fig:star_profiles} tests the rotating stellar models at the end of the relaxation. The left panel compares the azimuthal velocity of the SPH particles with the target profile
$v_\phi=\Omega(s)\,s$ as a function of the cylindrical radius $s$. The RR particles follow the uniform rotation out to the surface, and the DR particles follow the
Hachisu law, including its peak at $s=A$, with a scatter of a few per cent that is largest
near the peak. The right panel shows the virial sum $2E_{\rm spin}+2E_{\rm int}+E_{\rm grav}$,
which vanishes in equilibrium. It oscillates while the star adjusts to the added rotational
support, the oscillation is damped within a few tens of dynamical times, and the sum stays
at zero during the subsequent equilibrium test. Both rotating stars therefore enter the
encounter in rotational and virial equilibrium.
\section{AV accuracy and Spurious heating in differentially rotating star}
\label{appendixB}
\begin{figure}[tbp]
\centering
    \includegraphics[width=\textwidth]{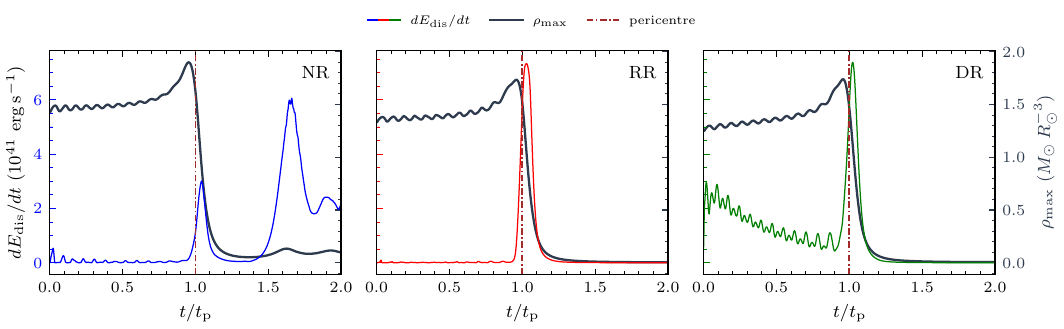}
    \caption{We plot evolution of the energy dissipation rate, $dE_{\rm dis}/dt$, and maximum density, $\rho_{\max}$, for the NR, RR, and DR stellar models during the strong-field encounter ($r_{\rm p}\simeq6.3\,r_{\rm g}$).  The vertical dash-dotted line indicates pericentre.}
\label{fig:AV_6rg}
\end{figure}

\begin{figure}[tbp]
\centering
    \includegraphics[width=\textwidth]{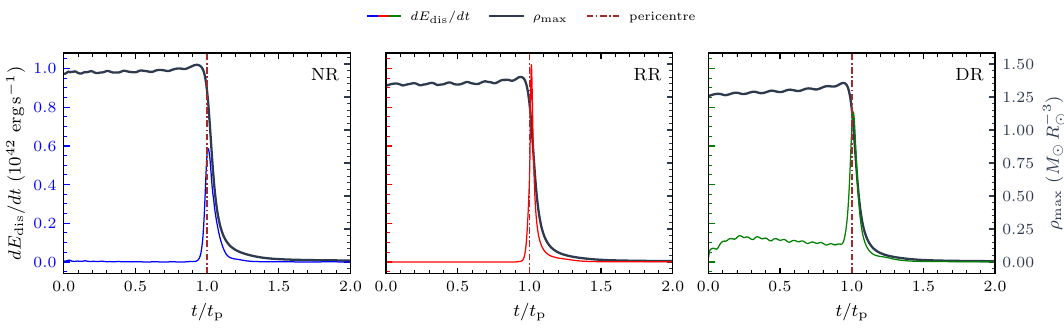}
    \caption{The energy dissipation rate, $dE_{\rm dis}/dt$, and maximum density, $\rho_{\max}$, are shown for the NR, RR, and DR stellar models undergoing the weak-field encounter at $r_{\rm p}\simeq25\,r_{\rm g}$. The vertical dash-dotted line marks pericentre passage.}
\label{fig:AV_25rg}
\end{figure}

Artificial viscosity (AV) is required to capture the shocks generated by tidal compression, but in a differentially rotating star it can also act on the background shear and produce spurious heating. Figures~\ref{fig:AV_6rg} and \ref{fig:AV_25rg} show the energy dissipation rate, $dE_{\rm dis}/dt$, and the maximum normalized central density, $\rho_{\max}$, for all three stellar models in both strong-field ($r_{\rm p}\simeq6.3\,r_{\rm g}$) and weak-field ($r_{\rm p}\simeq25\,r_{\rm g}$) encounters. Both sets of simulations correspond to near-polar retrograde orbits with stellar spin anti-aligned to the orbital angular momentum. Before the pericentre, the AV dissipation is negligible in the NR and RR stars. The DR model, however, exhibits enhanced dissipation due to its differential rotation. This contribution, however, remains comparatively smaller in the weak-field case ($r_{\rm p}\simeq25\,r_{\rm g}$) than in the strong-field encounter ($r_{\rm p}\simeq6.3\,r_{\rm g}$). 
Near the pericentre, the dissipation rate increases sharply in all three models, reaching peak values of order  as tidal compression drives shock formation. The maximum density $\rho_{\max}$ increases at the same time. The dissipation subsequently decreases as the shocked material expands and the debris is redistributed. The internal energy remains nearly constant during the approach to pericentre, at approximately $10^{48}\,\mathrm{erg}$, indicating that the pre-pericentre AV dissipation does not produce significant artificial heating of the stellar material.
The post-pericentre evolution also exhibits distinct secondary features depending on the encounter strength and stellar rotation profile. In the strong-field ($r_{\rm p}\simeq6.3\,r_{\rm g}$) NR simulation, a secondary dissipation peak appears around $2t_{\rm p}$, associated with the surviving bound core and the subsequent infall of debris, which collapses back onto the core. As the surviving bound core mass is much smaller for RR and DR models the peak is not observed at same simulation time. Overall, these results indicate that AV dissipation is dominated by physically relevant shock formation during tidal compression, while the additional dissipation associated with differential rotation remains subdominant and does not significantly modify the stellar structure prior to the encounter.

\begin{figure}
    \centering
    \begin{minipage}[b]{\textwidth}
        \centering
        \includegraphics[width=\textwidth]{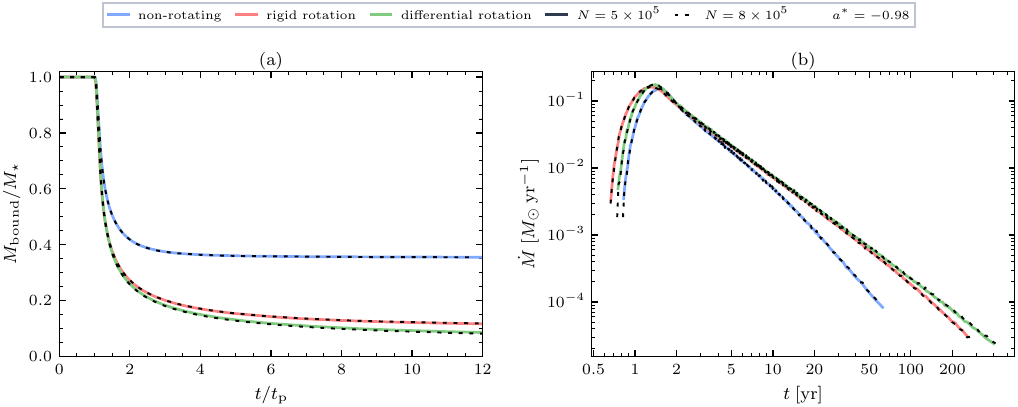}
    \end{minipage}
    
\caption{Resolution test for the partial disruption of a star on a near-polar orbit around a black hole with spin $a^* = -0.98$, for  non-rotating(blue), rigidly rotating (red) and a differentially rotating (green) star, at $N = 8\times10^5$ (dashed) and $N = 5\times10^5$ (solid) particles in aligned stellar spin configuration ($\theta_{\rm s}=0$). Left panel shows Bound stellar mass $M_\mathrm{bound}/M_\star$ as a function of time in units of the pericentre passage time $t_\mathrm{p}$. Right panel shows Mass fallback rate $\dot{M}$ as a function of time. The bound mass is converged to $<1$ per cent for the rigid rotator; for the differential rotator the difference grows to $\approx 5$ per cent by $10\,t_\mathrm{p}$. The peak fallback rate agrees to better than within 2 per cent in amplitude and 5 per cent in timing, and the decay phase agrees to within $\approx 2$ per cent up to $t \approx 100$\,yr and $\approx 4$ per cent (at most $10$ per cent at later times. }
\label{fig:convergence}
\end{figure}
\section{Duration of the strong tide at fixed penetration factor}
\label{app:fixed_beta}
The weak-field runs of Section~\ref{sec:25rg} use $\beta=0.75$ and the strong-field runs
use $\beta=0.555$, so a difference between the two sets could in principle come from $\beta$
rather than from relativity. A weak-field encounter with $\beta=0.555$ and
$r_{\rm p}\simeq25\,r_{\rm g}$ requires $M_{\rm BH}=6.25\times10^6\,M_\odot$. It lies at the
threshold for mass loss from a $\gamma=5/3$ polytrope, $\beta\simeq0.55$
\citep{GuillochonRamirezRuiz2013, Mainetti2017}, so it strips little or no mass and leaves
no core to compare with.

For a parabolic orbit the geodesic, measured in units of $GM_{\rm BH}/c^2$ and
$GM_{\rm BH}/c^3$, depends only on $r_{\rm p}/r_{\rm g}$, $a^*$ and $i$. At fixed
$r_{\rm p}/r_{\rm g}$ the black hole mass scales as $M_{\rm BH}\propto\beta^{-3/2}$, and so
does $GM_{\rm BH}/c^3$ in units of $t_\star$. The time spent inside $2r_{\rm p}$ therefore
separates into two factors,
\begin{equation}
\Delta t_{2{\rm p}}=f\,\frac{8\sqrt{2}}{3}\,\beta^{-3/2}\,t_\star ,
\label{eq:delta_t}
\end{equation}
where the second factor is the Newtonian value and $f(r_{\rm p}/r_{\rm g},a^*,i)$ contains
the relativistic corrections to the orbit and does not depend on $\beta$. We can find the factor $f$ by solving the numerical orbit equation as we did in Section~\ref{sec:pp}. Equation~\eqref{eq:delta_t} shows the contributions to the time spent near the strong-field zone by the test particle from $\beta$ and from relativistic corrections.

We integrated the geodesics as in Section~\ref{sec:pp} for
$M_{\rm BH}=6.25\times10^6\,M_\odot$, $\beta=0.555$, $a^*=0,\pm0.98$ and $i=90^\circ$,
 and $1^\circ$. The star stays inside $2r_{\rm p}$ for $9.9$--$10.2\,t_\star$,
against the Newtonian $9.1\,t_\star$, so $f=1.09$--$1.11$. The $\beta=0.75$ orbits at the
same radius give the same $f$ ($6.3$--$6.5\,t_\star$ against $5.8\,t_\star$), and in both
cases the orbit turns through $192^\circ$--$198^\circ$ inside $2r_{\rm p}$ and the
near-polar orbital plane tilts about $2^\circ$. At $6.3\,r_{\rm g}$ with the same
$\beta=0.555$, $f=1.42$--$1.66$ on the prograde, Schwarzschild and near-polar orbits and
rises to $2.5$ on the retrograde equatorial orbits. At fixed
$\beta$ the strong tide therefore lasts $1.3$--$2.2$ times longer at $6.3\,r_{\rm g}$ than
at $25\,r_{\rm g}$, which is a purely relativistic effect.

Between the two simulated sets, $\Delta t_{2{\rm p}}$ grows from $6.3$--$6.5\,t_\star$ to
$13$--$23\,t_\star$. A factor $(0.75/0.555)^{3/2}\simeq1.6$ of this comes from the lower
$\beta$ and the remaining factor $1.3$--$2.2$ from relativity. The lower $\beta$ cannot,
however, explain the loss of anti-aligned protection at $6.3\,r_{\rm g}$. At fixed
$r/r_{\rm p}$ the ratio of the Newtonian tidal acceleration to the self-gravity of the star
scales as $\beta^3$, so lowering $\beta$ from $0.75$ to $0.555$ weakens the tide by a factor
of $2.5$ while lengthening the stay by only $1.6$, and in Newtonian gravity brings the
encounter down to the threshold for mass loss. The strong stripping at $6.3\,r_{\rm g}$
(Section~\ref{sec:core}), and with it the long strong tide that strips the anti-aligned core
after it passes break-up (Section~\ref{sec:25rg_spin}), are therefore strong-field effects,
consistent with the interpretation of Section~\ref{sec:25rg}.

\section{Convergence tests}
\label{app:convergence}
To test numerical convergence, we repeated the near-polar and aligned spin configuration with black hole spin $a^*=-0.98$ encounters at $r_{\rm p}\simeq6.3\,r_{\rm g}$ for all three stellar spin models at two resolutions, $N = 5\times10^5$ and $N = 8\times10^5$ particles. Figure~\ref{fig:convergence} shows that fallback rates agree closely at both resolutions upto two decimal places in fallback rate, $\dot M$. The peak fallback, $\dot M_{\rm peak}\simeq0.15$--$0.17\,M_\odot\,{\rm yr}^{-1}$ at  $t_{\rm peak}\simeq1.3$--$1.5$\,yr, is reproduced. The power-law decay agrees out to $t\geq100$\,yr, where the two resolutions differ by no more than a few per cent. The evolution of the bound stellar mass is converged to within the line width throughout: the low and high resolution curves are indistinguishable from the onset of stripping at $t=t_{\rm p}$ through the asymptotic bound fractions, $M_{\rm bound}/M_\star=0.36$, $0.12$ and $0.08$ for NR, RR and DR at $t = 12\,t_{\rm p}$. Resolution changes our results far less than the stellar rotation profile does, so the differences between rigid and differential rotation are physical rather than numerical.


\begin{thebibliography}{}
\bibitem[Andalman et al.(2022)]{Andalman2022} Andalman, Z.~L., Liska, M.~T.~P., Tchekhovskoy, A., Coughlin, E.~R., \& Stone, N. 2022, \mnras, 510, 1627. \href{https://doi.org/10.1093/mnras/stab3444}{doi:10.1093/mnras/stab3444}

\bibitem[Ayal et al.(2000)]{Ayal2000} Ayal, S., Livio, M., \& Piran, T. 2000, \apj, 545, 772. \href{https://doi.org/10.1086/317835}{doi:10.1086/317835}

\bibitem[Bandopadhyay et al.(2026)]{Bandopadhyay2026}
Bandopadhyay, A., Amend, B., Coughlin, E.~R., Nixon, C.~J.,
Pasham, D.~R., \& Wevers, T. 2026, \apj, 1007, 181.
\href{https://doi.org/10.3847/1538-4357/ae8f31}{doi:10.3847/1538-4357/ae8f31}
[\href{https://arxiv.org/abs/2606.02692}{arXiv:2606.02692}]

\bibitem[Balsara(1995)]{Balsara1995} Balsara, D.~S. 1995, J. Comput. Phys., 121, 357
\bibitem[Banerjee et al.(2023)]{Banerjee2023} Banerjee, P., Garain, D., Chowdhury, S., Singh, D., Joshi, R., \& Sarkar, T. 2023, \mnras, 522, 4332. \href{https://doi.org/10.1093/mnras/stad1284}{doi:10.1093/mnras/stad1284} [\href{https://arxiv.org/abs/2212.09122}{arXiv:2212.09122}]

\bibitem[Banerjee et al.(2026a)]{Banerjee2026a} Banerjee, P., Chakraborty, K., Mondal, N., \& Sarkar, T. 2026a, arXiv:2603.26442 [\href{https://arxiv.org/abs/2603.26442}{arXiv:2603.26442}]

\bibitem[Banerjee et al.(2026b)]{Banerjee2026b} Banerjee, P., Chakraborty, K., Mondal, N., \& Sarkar, T. 2026b, arXiv:2610.27775 [\href{https://arxiv.org/abs/2610.27775}{arXiv:2610.27775}]

\bibitem[Bonnerot et al.(2016)]{Bonnerot2016} Bonnerot, C., Rossi, E.~M., Lodato, G., \& Price, D.~J. 2016, \mnras, 455, 2253. \href{https://doi.org/10.1093/mnras/stv2411}{doi:10.1093/mnras/stv2411}

\bibitem[Calder{\'o}n et al.(2026)]{Calderon2026} Calder{\'o}n, D., Pejcha, O., Metzger, B.~D., Duffell, P.~C., \& Rosswog, S. 2026,  Astron. Nachr., 347, e70092. \href{https://doi.org/10.1002/asna.70092}{doi:10.1002/asna.70092}, [\href{https://arxiv.org/abs/2603.10124}{arXiv:2603.10124}]

\bibitem[Cannizzo et al.(1990)]{Cannizzo1990} Cannizzo, J.~K., Lee, H.~M., \& Goodman, J. 1990, \apj, 351, 38. \href{https://doi.org/10.1086/168442}{doi:10.1086/168442}

\bibitem[Carter \& Luminet(1983)]{CarterLuminet1983} Carter, B. \& Luminet, J.-P. 1983, \aap, 121, 97

\bibitem[Chan et al.(2026)]{Chan2026} Chan, H.-S., Ryu, T., Krolik, J., \& Piran, T. 2026, \apj, 1003, 221. \href{https://doi.org/10.3847/1538-4357/ae67fa}{doi:10.3847/1538-4357/ae67fa}

\bibitem[Cheng \& Evans(2013)]{ChengEvans2013} Cheng, R.~M. \& Evans, C.~R. 2013, \prd, 87, 104010. \href{https://doi.org/10.1103/PhysRevD.87.104010}{doi:10.1103/PhysRevD.87.104010}

\bibitem[Coughlin \& Nixon(2019)]{CoughlinNixon2019} Coughlin, E.~R. \& Nixon, C.~J. 2019, \apjl, 883, L17. \href{https://doi.org/10.3847/2041-8213/ab412d}{doi:10.3847/2041-8213/ab412d}

\bibitem[Dai et al.(2015)]{Dai2015} Dai, L., McKinney, J.~C., \& Miller, M.~C. 2015, \apjl, 812, L39. \href{https://doi.org/10.1088/2041-8205/812/2/L39}{doi:10.1088/2041-8205/812/2/L39}

\bibitem[Dehnen \& Aly(2012)]{DehnenAly2012} Dehnen, W. \& Aly, H. 2012, \mnras, 425, 1068

\bibitem[Evans \& Kochanek(1989)]{EvansKochanek1989} Evans, C.~R. \& Kochanek, C.~S. 1989, \apjl, 346, L13. \href{https://doi.org/10.1086/185567}{doi:10.1086/185567}

\bibitem[Franchini et al.(2016)]{Franchini2016} Franchini, A., Lodato, G., \& Facchini, S. 2016, \mnras, 455, 1946. \href{https://doi.org/10.1093/mnras/stv2417}{doi:10.1093/mnras/stv2417}

\bibitem[Frank \& Rees(1976)]{FrankRees} Frank, J. \& Rees, M.~J. 1976, \mnras, 176, 633. \href{https://doi.org/10.1093/mnras/176.3.633}{doi:10.1093/mnras/176.3.633}

\bibitem[Gafton \& Rosswog(2019)]{GaftonRosswog2019} Gafton, E. \& Rosswog, S. 2019, \mnras, 487, 4790. \href{https://doi.org/10.1093/mnras/stz1530}{doi:10.1093/mnras/stz1530}

\bibitem[Gafton et al.(2015)]{Gafton2015} Gafton, E., Tejeda, E., Guillochon, J., Korobkin, O., \& Rosswog, S. 2015, \mnras, 449, 771. \href{https://doi.org/10.1093/mnras/stv350}{doi:10.1093/mnras/stv350}

\bibitem[Garain \& Sarkar(2024)]{GarainSarkar2024} Garain, D., \& Sarkar, T. 2024, \apj, 967, 167. \href{https://doi.org/10.3847/1538-4357/ad3dfa}{doi:10.3847/1538-4357/ad3dfa}

\bibitem[Garain \& Sarkar(2025)]{GarainSarkar2025} Garain, D., \& Sarkar, T. 2025, \mnras, 542, 839. \href{https://doi.org/10.1093/mnras/staf1270}{doi:10.1093/mnras/staf1270}

\bibitem[Golightly et al.(2019)]{Golightly2019} Golightly, E.~C.~A., Coughlin, E.~R., \& Nixon, C.~J. 2019, \apj, 872, 163. \href{https://doi.org/10.3847/1538-4357/aafd2f}{doi:10.3847/1538-4357/aafd2f} [\href{https://arxiv.org/abs/1901.03717}{arXiv:1901.03717}]

\bibitem[Guillochon \& Ramirez-Ruiz(2013)]{GuillochonRamirezRuiz2013} Guillochon, J. \& Ramirez-Ruiz, E. 2013, \apj, 767, 25. \href{https://doi.org/10.1088/0004-637X/767/1/25}{doi:10.1088/0004-637X/767/1/25}

\bibitem[Hayasaki et al.(2013)]{Hayasaki2013} Hayasaki, K., Stone, N., \& Loeb, A. 2013, \mnras, 434, 909. \href{https://doi.org/10.1093/mnras/stt871}{doi:10.1093/mnras/stt871}

\bibitem[Hayasaki et al.(2016)]{Hayasaki2016} Hayasaki, K., Stone, N., \& Loeb, A. 2016, \mnras, 461, 3760. \href{https://doi.org/10.1093/mnras/stw1387}{doi:10.1093/mnras/stw1387}

\bibitem[Hills(1975)]{Hills1975} Hills, J.~G. 1975, \nat, 254, 295. \href{https://doi.org/10.1038/254295a0}{doi:10.1038/254295a0}

\bibitem[Kagaya et al.(2019)]{Kagaya2019} Kagaya, K., Yoshida, S., \& Tanikawa, A. 2019, arXiv:1901.05644 [\href{https://arxiv.org/abs/1901.05644}{arXiv:1901.05644}]

\bibitem[Kesden(2012)]{Kesden2012} Kesden, M. 2012, \prd, 85, 024037. \href{https://doi.org/10.1103/PhysRevD.85.024037}{doi:10.1103/PhysRevD.85.024037}

\bibitem[Kochanek(1994)]{Kochanek1994} Kochanek, C.~S. 1994, \apj, 422, 508. \href{https://doi.org/10.1086/173745}{doi:10.1086/173745}

\bibitem[Komatsu et al.(1989)]{Komatsu1989} Komatsu, H., Eriguchi, Y., \& Hachisu, I. 1989, \mnras, 237, 355. \href{https://doi.org/10.1093/mnras/237.2.355}{doi:10.1093/mnras/237.2.355}

\bibitem[Laguna et al.(1993)]{Laguna1993} Laguna, P., Miller, W.~A., Zurek, W.~H., \& Davies, M.~B. 1993, \apjl, 410, L83. \href{https://doi.org/10.1086/186885}{doi:10.1086/186885}

\bibitem[Law-Smith et al.(2020)]{LawSmith2020} Law-Smith, J.~A.~P., Coulter, D.~A., Guillochon, J., Mockler, B., \& Ramirez-Ruiz, E. 2020, \apj, 905, 141. \href{https://doi.org/10.3847/1538-4357/abc489}{doi:10.3847/1538-4357/abc489}

\bibitem[Leloudas et al.(2016)]{Leloudas2016} Leloudas, G., Fraser, M., Stone, N.~C., et al. 2016, Nature Astronomy, 1, 0002. \href{https://doi.org/10.1038/s41550-016-0002}{doi:10.1038/s41550-016-0002}

\bibitem[Li et al.(2021)]{Li2021}
Li, J., Lai, D., Anderson, K.~R., \& Pu, B. 2021, \mnras, 501, 1621. \href{https://doi.org/10.1093/mnras/staa3779}{doi:10.1093/mnras/staa3779}

\bibitem[Liptai \& Price(2019)]{LiptaiPrice2019} Liptai, D. \& Price, D.~J. 2019, \mnras, 485, 819. \href{https://doi.org/10.1093/mnras/stz111}{doi:10.1093/mnras/stz111} [\href{https://arxiv.org/abs/1901.08064}{arXiv:1901.08064}]

\bibitem[Liptai et al.(2019)]{Liptai2019} Liptai, D., Price, D.~J., Mandel, I., \& Lodato, G. 2019, arXiv:1910.10154 [\href{https://arxiv.org/abs/1910.10154}{arXiv:1910.10154}]

\bibitem[Lodato et al.(2009)]{Lodato2009} Lodato, G., King, A.~R., \& Pringle, J.~E. 2009, \mnras, 392, 332. \href{https://doi.org/10.1111/j.1365-2966.2008.14049.x}{doi:10.1111/j.1365-2966.2008.14049.x}


\bibitem[Lodato \& Rossi(2011)]{LodatoRossi2011} Lodato, G. \& Rossi, E.~M. 2011, \mnras, 410, 359. \href{https://doi.org/10.1111/j.1365-2966.2010.17448.x}{doi:10.1111/j.1365-2966.2010.17448.x} [\href{https://arxiv.org/abs/1008.4589}{arXiv:1008.4589}]


\bibitem[Mahapatra et al.(2026a)]{Mahapatra2026a} Mahapatra, A., Pandey, A., Garain, D., \& Sarkar, T. 2026, JCAP, 01, 010. \href{https://doi.org/10.1088/1475-7516/2026/01/010}{doi:10.1088/1475-7516/2026/01/010} [\href{https://arxiv.org/abs/2410.12727}{arXiv:2410.12727}]

\bibitem[Mahapatra et al.(2026b)]{Mahapatra2026b} Mahapatra, A., Pandey, A., Banerjee, P., \& Sarkar, T. 2026, \apj, 1004, 167. \href{https://doi.org/10.3847/1538-4357/ae6e3e}{doi:10.3847/1538-4357/ae6e3e}

\bibitem[Mahapatra et al.(2026c)]{Mahapatra2026c}
Mahapatra, A., Pandey, A., Banerjee, P., \& Sarkar, T. 2026, \apj, 1011, 266.
\href{https://doi.org/10.3847/1538-4357/aeaccb}{doi:10.3847/1538-4357/aeaccb}
[\href{https://arxiv.org/abs/2601.04870}{arXiv:2601.04870}]

\bibitem[Mainetti et al.(2017)]{Mainetti2017} Mainetti, D., Lupi, A., Campana, S., et al. 2017, \aap, 600, A124. \href{https://doi.org/10.1051/0004-6361/201630092}{doi:10.1051/0004-6361/201630092}

\bibitem[Mummery et al.(2024)]{Mummery2024} Mummery, A., van Velzen, S., Nathan, E., Ingram, A., Hammerstein, E., Fraser-Taliente, L., \& Balbus, S. 2024, \mnras, 527, 2452. \href{https://doi.org/10.1093/mnras/stad3001}{doi:10.1093/mnras/stad3001}

\bibitem[Mummery \& van Velzen(2025)]{Mummery2025}
Mummery, A., \& van Velzen, S. 2025, \mnras, 541, 429.
\href{https://doi.org/10.1093/mnras/staf938}{doi:10.1093/mnras/staf938}

\bibitem[Mummery(2026)]{Mummery2026} Mummery, A. 2026, arXiv:2601.14483 [\href{https://arxiv.org/abs/2601.14483}{arXiv:2601.14483}]

\bibitem[Nixon et al.(2021)]{Nixon2021} Nixon, C.~J., Coughlin, E.~R., \& Miles, P.~R. 2021, \apj, 922, 168. \href{https://doi.org/10.3847/1538-4357/ac1bb8}{doi:10.3847/1538-4357/ac1bb8} [\href{https://arxiv.org/abs/2108.04242}{arXiv:2108.04242}]

\bibitem[Nolthenius \& Katz(1982)]{Nolthenius1982} Nolthenius, R.~A. \& Katz, J.~I. 1982, \apj, 263, 377. \href{https://doi.org/10.1086/160511}{doi:10.1086/160511}

\bibitem[Phinney(1989)]{Phinney1989} Phinney, E.~S. 1989, in IAU Symp.\ 136, The Center of the Galaxy, ed.\ M.~Morris (Dordrecht: Kluwer), 543

\bibitem[Piran et al.(2026)]{Piran2026}
Piran, T., Krolik, J., \& Ryu, T. 2026, \apj, 1001, 48.
\href{https://doi.org/10.3847/1538-4357/ae50ff}{doi:10.3847/1538-4357/ae50ff}

\bibitem[Price et al.(2024)]{Price2024}
Price, D.~J., Liptai, D., Mandel, I., et al. 2024, \apjl, 971, L46.
\href{https://doi.org/10.3847/2041-8213/ad6862}{doi:10.3847/2041-8213/ad6862}
[\href{https://arxiv.org/abs/2404.09381}{arXiv:2404.09381}]


\bibitem[Pringle(1981)]{Pringle1981} Pringle, J.~E. 1981, \araa, 19, 137. \href{https://doi.org/10.1146/annurev.aa.19.090181.001033}{doi:10.1146/annurev.aa.19.090181.001033}


\bibitem[Rees(1988)]{Rees} Rees, M.~J. 1988, \nat, 333, 523. \href{https://doi.org/10.1038/333523a0}{doi:10.1038/333523a0}

\bibitem[Ryu et al.(2020a)]{Ryu2020I} Ryu, T., Krolik, J., Piran, T., \& Noble, S.~C. 2020a, \apj, 904, 98. \href{https://doi.org/10.3847/1538-4357/abb3cf}{doi:10.3847/1538-4357/abb3cf}
\bibitem[Ryu et al.(2020b)]{Ryu2020II} Ryu, T., Krolik, J., Piran, T., \& Noble, S.~C. 2020b, \apj, 904, 99. \href{https://doi.org/10.3847/1538-4357/abb3cd}{doi:10.3847/1538-4357/abb3cd} [\href{https://arxiv.org/abs/2001.03502}{arXiv:2001.03502}]

\bibitem[Ryu et al.(2020c)]{Ryu2020III} Ryu, T., Krolik, J., Piran, T., \& Noble, S.~C. 2020c, \apj, 904, 100. \href{https://doi.org/10.3847/1538-4357/abb3ce}{doi:10.3847/1538-4357/abb3ce}
\bibitem[Ryu et al.(2020d)]{Ryu2020IV} Ryu, T., Krolik, J., Piran, T., \& Noble, S.~C. 2020d, \apj, 904, 101. \href{https://doi.org/10.3847/1538-4357/abb3cc}{doi:10.3847/1538-4357/abb3cc}

\bibitem[Sacchi \& Lodato(2019)]{SacchiLodato2019} Sacchi, A. \& Lodato, G. 2019, \mnras, 486, 1833. \href{https://doi.org/10.1093/mnras/stz981}{doi:10.1093/mnras/stz981} [\href{https://arxiv.org/abs/1904.02424}{arXiv:1904.02424}]


\bibitem[S{\c a}dowski et al.(2016)]{Sadowski2016} S{\c a}dowski, A., Tejeda, E., Gafton, E., Rosswog, S., \& Abarca, D. 2016, \mnras, 458, 4250. \href{https://doi.org/10.1093/mnras/stw589}{doi:10.1093/mnras/stw589}

\bibitem[Shakura \& Sunyaev(1973)]{ShakuraSunyaev1973} Shakura, N.~I. \& Sunyaev, R.~A. 1973, \aap, 24, 337

\bibitem[Shen \& Matzner(2014)]{ShenMatzner2014} Shen, R.-F. \& Matzner, C.~D. 2014, \apj, 784, 87. \href{https://doi.org/10.1088/0004-637X/784/2/87}{doi:10.1088/0004-637X/784/2/87} [\href{https://arxiv.org/abs/1305.5570}{arXiv:1305.5570}]


\bibitem[Shiokawa et al.(2015)]{ShiokawaEtAl2015} Shiokawa, H., Krolik, J.~H., Cheng, R.~M., Piran, T., \& Noble, S.~C. 2015, \apj, 804, 85. \href{https://doi.org/10.1088/0004-637X/804/2/85}{doi:10.1088/0004-637X/804/2/85}

\bibitem[Stone \& Loeb(2012)]{StoneLoeb2012} Stone, N. \& Loeb, A. 2012, \prl, 108, 061302. \href{https://doi.org/10.1103/PhysRevLett.108.061302}{doi:10.1103/PhysRevLett.108.061302}

\bibitem[Stone et al.(2013)]{StoneSariLoeb2013} Stone, N., Sari, R., \& Loeb, A. 2013, \mnras, 435, 1809. \href{https://doi.org/10.1093/mnras/stt1270}{doi:10.1093/mnras/stt1270}

\bibitem[Tejeda et al.(2017)]{Tejeda2017} Tejeda, E., Gafton, E., Rosswog, S., \& Miller, J.~C. 2017, \mnras, 469, 4483. \href{https://doi.org/10.1093/mnras/stx1089}{doi:10.1093/mnras/stx1089} [\href{https://arxiv.org/abs/1701.00303}{arXiv:1701.00303}]


\bibitem[van Velzen et al.(2019)]{vanVelzen2019} van Velzen, S., Stone, N.~C., Metzger, B.~D., et al. 2019, \apj, 878, 82. \href{https://doi.org/10.3847/1538-4357/ab1844}{doi:10.3847/1538-4357/ab1844} [\href{https://arxiv.org/abs/1809.00003}{arXiv:1809.00003}]


\bibitem[Wendland(1995)]{Wendland1995} Wendland, H. 1995, Adv. Comput. Math., 4, 389

\bibitem[Wang(2024)]{Wang2024GSN} Wang, D. 2024, \aap, 682, L14. \href{https://doi.org/10.1051/0004-6361/202348613}{doi:10.1051/0004-6361/202348613}
\end{thebibliography}
\end{document}